%% file: LSMT.tex
\documentclass[12pt,a4paper]{article}
\usepackage[utf8]{inputenc}
\usepackage{physics,amsmath,bm}
\usepackage{amsfonts}
\usepackage{amssymb}
\usepackage[left=.14\paperwidth,right=.14\paperwidth,top=.14\paperheight,bottom=.14\paperheight]{geometry}
\usepackage{cite}

\usepackage{slashed}
\usepackage{hyperref}
\usepackage{graphicx}
\usepackage{caption}
\usepackage{float}
\usepackage[labelformat=simple]{subcaption}
\usepackage{authblk}
\hypersetup{
	unicode=true,          
	pdftoolbar=true,        
	pdfmenubar=true,        
	pdffitwindow=false,     
	pdfstartview={FitH},    
	pdftitle={LocalisedSMatrix},    
	pdfauthor={},     
	pdfsubject={},   
	pdfcreator={},   
	pdfproducer={DK,AP}, 
	pdfkeywords={S-Matrix} {localisation effects} {finite-volume effects}, 
	pdfnewwindow=true,      
	colorlinks=true,       
	linkcolor=blue,          
	citecolor=red,        
	filecolor=black,      
	urlcolor=blue           
}

\include{macros}

\renewcommand{\theequation}{\arabic{section}.\arabic{equation}}

\definecolor{ao}{rgb}{0.0, 0.0, 1.0}
\newenvironment{bl}[1]{{\color{ao}  #1}}

\newcommand*{\email}[1]{%
	\footnotesize\href{mailto:#1}{\bl{#1}}
}

\title{\bf {Towards a Localised S-Matrix Theory}}
\author[ ]{{\large Dimitrios Karamitros}$^{\:a}\,$\footnote{\email{Dimitrios.Karamitros@manchester.ac.uk}} \hspace{-11pt}}
\author[ ]{{\large Apostolos Pilaftsis}$^{\:a,b}\,$\footnote{\email{Apostolos.Pilaftsis@manchester.ac.uk}}}

\affil[ ]{\small \it \hspace{1cm}$^a$Department of Physics and Astronomy, University of Manchester,\newline Manchester, M13
	9PL, United Kingdom \vspace*{0.25cm}}
\affil[ ]{ \small \it $^b$Theoretical Physics Department, CERN, CH-1211 Geneva 23, Switzerland}

\date{}

\begin{document}
{
	\vspace*{-1.5cm}
	\begin{flushright}
		CERN-TH-2022-138\\
		August 2022
	\end{flushright}
}	

\renewcommand\abstractname{\textsc{ABSTRACT}}

{\let\newpage\relax\maketitle}

\maketitle

\begin{abstract}
\noindent
    We formulate an S-matrix theory in which localisation effects of the particle interactions involved in a scattering process are consistently taken into account\-. In the limit of an infinite spread of all interactions, the S-matrix assumes its standard form. To better understand the significance of the emerging quantum phenomena in this formalism, we consider a solvable field-theoretic model with spatial Gaussian spreads at the interaction vertices. This solvable model, which was previously introduced in the literature,  enables accurate descriptions of detection regions that are either close to or far from the source. In close analogy with light diffraction in classical optics, we call these two regions near-field and far-field zones, or the Fresnel and Fraunhofer regions. We revisit the question whether mixed mediators produce an oscillating pattern if their detection occurs in the Fresnel region. Besides\- corroborating certain earlier findings of the S-matrix amplitude in the forward Fresnel and Fraunhofer regimes, we observe several novel features with respect to its angular dependence which have not been accounted before in the literature. In particular, we obtain a ``quantum obliquity factor'' that suppresses particle propagation in the backwards direction, thereby providing an explicit quantum field-theoretic description for its origin in diffractive optics. Present and future colliders, as well as both short and long baseline neutrino experiments, would greatly benefit from the many predictions that can be offered from such a holistic localised S-matrix theory.
\end{abstract}

ArXiv ePrint: \href{https://arxiv.org/abs/2208.10425}{\bl{2208.10425}}

\newpage

 \tableofcontents

\newpage

\section{Introduction}\label{sec:intro}
\setcounter{equation}{0}

Despite the initial scepticism expressed by Einstein, Podolsky and Rosen (EPR)~\cite{Einstein:1935rr} concerning the completeness  of Quantum Mechanics (QM), the non-local nature of the quantum-mechanical wavefunction has  been vindicated by  now  in vast number of experiments  through the violation of the famous Bell's inequalities~\cite{Bell:1964kc}. An astounding physical consequence of the property of non-locality in QM is the emergence of the so-called quantum entanglement between quantum states which became the primary engineering principle in many applications of modern quantum theory, including quantum information, quantum technology and particle physics~\cite{Feynman:1981tf,Alonso:2022oot}.

With the advent of the more complete framework of Quantum Field Theory~(QFT), physical observables  associated with scattering processes are encoded in the so-called S-matrix~\cite{Lehmann:1954rq, Eden:1966dnq}. The development of a unitary S-matrix theory allowed us to make accurate predictions for (differential) cross sections of $2\to n$ processes in momentum space which have been tested with great success  in collider experiments (for a review, see~\cite{Workman:2022ynf}). However, in its standard formulation, the S-matrix provides no space-time information of the non-local form of Feynman propagators, thus limiting considerably its degree of applicability. For instance, the production and decay of long-lived particles, like those that occur in $K$-, $B$- and $D$-meson systems, would necessitate the knowledge of the production and detection vertices, along with the momenta of the particles in both the initial and final state of such processes. Likewise, the necessity of describing the observed phenomenon of neutrino oscillations in space within the framework of QFT~\cite{Giunti:1993se,Grimus:1996av,Campagne:1997fu,Kiers:1997pe,Ioannisian:1998ch,Cardall:1999bz,Beuthe:2001rc,Kopp:2009fa,Akhmedov:2009rb,Akhmedov:2010ms,Akhmedov:2010ua,Naumov:2013bea,Naumov:2020yyv,Grimus:2019hlq,Cheng:2022lys,Naumov:2022kwz} would require the development of an S-matrix theory that takes into account finite-size localisation effects of particle interactions. Hereafter, we refer to such a theory, for brevity, as a Localised S-Matrix Theory (LSMT). 

Another important application of such an LSMT will be to successfully regulate $t$-channel kinematic singularities of tree-level transition amplitudes~\cite{Peierls:1961zz,Coleman:1965xm,Nowakowski:1993iu,Ginzburg:1995bc} that appear in the physical region of the phase space. In the LSMT framework, this can be done without appealing,  for example, to statistical uncertainties of the particle momenta in the colliding muon beams~\cite{Kotkin:1992bj,Melnikov:1996na,Melnikov:1996iu,Grzadkowski:2021kgi}. Hence, the dynamics regulating such $t$-channel singularities can be fundamentally different from that presented in other approaches. Other applications of an LSMT may include new-physics searches for displaced vertices during the hadronization process at high-energy colliders like~LHC~\cite{LHCb:2014osd,Bondarenko:2019tss}, or its peripheral experiments~\cite{Beacham:2019nyx}, FASER~\cite{Feng:2017uoz,FASER:2018eoc}, MATHUSLA~\cite{Chou:2016lxi,Curtin:2018mvb} and SHiP~\cite{Alekhin:2015byh}. Such considerations may lead to an improved interpretation of the experimental~data.

In this paper we aim to formulate an S-matrix theory in which  effects in particle interactions are consistently taken into account in scattering processes. The proposed construction of the LSMT receives its standard S-matrix form in the well defined limit in which the spread of all interaction vertices is taken to be infinite. To illustrate the key features of our localised S-matrix formalism, we will consider a $2\to 2$ scattering process within a solvable QFT model in which the production and detection vertices are assumed to have a spatial spread of Gaussian form. 

The aforementioned solvable QFT model was previously introduced in~\cite{Ioannisian:1998ch} where several basic  properties of propagation and oscillation of neutrinos were analysed in two physical regions that depend on the distance $\absl$ of the detector from the source. Here, we further consolidate these earlier findings by borrowing a terminology known from light diffraction in classical optics. Exactly as~in diffractive optics, depending on $\absl$, we have two regions which we call the near-field and far-field zones, or the Fresnel and Fraunhofer regions. These two regions depend on the spatial  spread of the production or detection vertices, which we generically denote as $\dl$, and the magnitude~$\absp$ of the net three-momentum ${\bf p}$ of all particles in the initial or final state. Hence, the Fresnel (near-field) zone refers typically to distances $\absl$ in the interval, ${0 \le \absl \lesssim \absp\,\dl^2}$, whilst the Fraunhofer (far-field) regime sets on in its full glory when~${\absl \gg  \absp\, \dl^2}$. 

In this article we also study in more detail all emerging quantum phenomena that result from our localised S-matrix formalism in the context of the solvable QFT model presented in~\cite{Ioannisian:1998ch}. In particular, we re-examine the question whether mediators of a particle-mixing system like neutrinos produce an oscillating pattern their detection occurs in the Fresnel region. As well as confirming certain earlier results~\cite{Ioannisian:1998ch} concerning the analytic behaviour of the S-matrix amplitude in the forward Fresnel and Fraunhofer zones, we find several novel features with respect to its angular dependence which have not been discussed in adequate detail before in the literature. Most remarkably, we obtain a ``quantum obliquity factor'' in the transition amplitude that suppresses the propagation of the mediator in the backwards direction, when the latter has real momentum. This suppression is achieved without imposing the restrictions owing to the Huygens--Fresnel's principle, but it is rather a consequence of the inherent boundary conditions that the Feynman propagator obeys. Thus, an alternative quantum field-theoretic explanation can be obtained for the origin of the obliquity factor in diffractive optics~\cite{Optics_1999}. 

The localised S-matrix theory that we will be developing here could be utilised at high-energy colliders to describe hadronization in a framework consistent with quantum mechanics, beyond the so-called Lund model~\cite{Andersson:1983ia,Ferreres-Sole:2018vgo}. Likewise, short and long baseline neutrino experiments would benefit from the development of LSMT that will provide a more accurate interpretation of the low-energy neutrino oscillation data.  

The paper is organised as follows. After this introductory section, we briefly review in Section~\ref{sec:non_local_intro} the basic results emanating from the conventional S-matrix theory by considering a $2\to2$ process in a simple scalar field theory. We then discuss a localised modification of this standard S-matrix theory, and present exact analytic results within a solvable QFT model. 
In close analogy with diffractive optics, we present in Section~\ref{sec:approx} approximate analytic expressions of the S-matrix amplitude in the near- and far-field zones as functions of the distance $\absl$ of the detector from the source.
In Section~\ref{sec:numerics} we give exact results by analysing numerical examples, which confirm explicitly the validity of the Fresnel and Fraunhofer approximations discussed in the previous section. In addition, we show the complete angular dependence of the transition amplitude in the polar coordinates $(\absl,\theta)$, where $\theta$ (with $0\le \theta \le \pi)$ is the angle between the distance vector~$\vecl$ and the total three-momentum~${\bf p}$ of the colliding particles in the initial state of the scattering process. 
Finally, Section~\ref{sec:summary} provides a succinct summary of our results and discusses possible future research directions. Technical details of the calculation of the localised S-matrix amplitude within a solvable QFT model are given in Appendices~\ref{app:amplitude}, \ref{app:angular} and~\ref{app:radial}.

\section{The Localised S-Matrix}\label{sec:non_local_intro}
\setcounter{equation}{0}

In a local QFT, the notion of non-locality enters through the Feynman propagator, which we denote here as~$\Delta_{\rm F} (x,y)$. A remarkable property of the Feynman propagator is that it has non-zero support for two space-time points, $x$ and $y$, which happen to be localised at space-like separations, i.e.~$\Delta_{\rm F} (x,y) \neq 0$, for $(x-y)^2 < 0$. In~fact, this property encodes the counter-intuitive non-local phenomenon of quantum entanglement in QM which was called by Einstein in a letter to Max Born in 1947: ``spooky action at a distance''. But exactly as happens with quantum entanglement in QM, no true information between any two space-like separated points,~$x$ and~$y$, can be transferred faster than the speed of light, and as such, QFT respects causality~\cite{Maiani:1994zi,Eden:1966dnq,Donoghue:2020mdd}.

In the standard S-matrix theory emerging from QFT~\cite{Eden:1966dnq,Pokorski:1987ed,Weinberg:1995mt,Peskin:1995ev,Srednicki:2007qs,Schwartz:2014sze}, the transition amplitudes resulting from the so-called Lehmann--Symanzik--Zimmermann (LSZ) formalism~\cite{Lehmann:1954rq} do not depend on space-time, but only on the four-momenta of all particles in the initial and final state of a scattering process. Hence, any information concerning the location of the interactions in a $2\to n$ process is lost after integration over an infinite space-time volume. If these interactions are restricted locally within a space-time volume of finite size, the resulting S-matrix will depend on the space-time coordinates and other quantities that parameterise the spread due to coherent QM uncertainties at the interaction vertices. The formulation of such a Localised S-Matrix Theory (LSMT) is that we wish to put forward in this paper$\,$\footnote{For other attempts along this research direction, see~\cite{Giunti:1993se,Grimus:1996av,Campagne:1997fu,Kiers:1997pe,Cardall:1999bz,Beuthe:2001rc,Kopp:2009fa,Akhmedov:2009rb,Akhmedov:2010ms,Akhmedov:2010ua,Naumov:2013bea,Naumov:2020yyv,Grimus:2019hlq,Cheng:2022lys,Naumov:2022kwz,Nowakowski:1993iu,Ginzburg:1995bc,Kotkin:1992bj,Melnikov:1996iu,Melnikov:1996na,Grzadkowski:2021kgi}.}.

In the remainder of the section, we consider a $2\to 2$ scattering process in a (local) QFT model with scalar fields. We first recall the simple derivation of the ordinary S-matrix element for such a process in the Born approximation. We then turn our attention to the localisation profiles introduced in this S-matrix element, when the interaction vertices occur in a confined region of space-time. Finally, we revisit the analytic results obtained in the solvable QFT model of~\cite{Ioannisian:1998ch}.

\subsection{Standard S-Matrix Theory}

Let us consider a simple scalar field theory consisting of five real scalar fields: $S_{1,2}$, $\chi_{1,2}$, and~$\Phi$. 
The interactions of this QFT model are governed by the local Lagrangian,  
\begin{eqnarray}
	{\cal L}_{\rm int} (x)\, =\, \lambda  \ S_1(x) \chi_1(x)\, \Phi(x)\, +\, g  \ S_2(x) \chi_2(x)\, \Phi(x) \;,
	\label{eq:Lint_local}
\end{eqnarray}
where $\lambda$ and $g$ are two real couplings. In the Born approximation of this QFT model, any scattering between $S_{1,2}$ and $\chi_{1,2}$ will be mediated by the exchange of a particle $\Phi$ involving a single Feynman diagram.

\begin{figure}[t!]
	\centering\includegraphics[width=0.6\textwidth ]{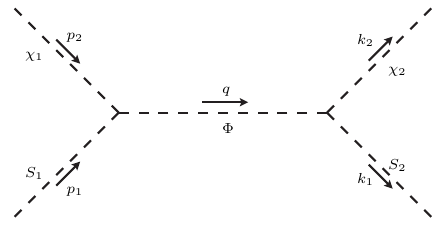}
	\caption{The Feynman diagram for the process $S_1(p_1) \ \chi_1(p_2)  \to S_2(k_1) \  \chi_2(k_2)$.}
	\label{fig:local_FeynD}
\end{figure}

To set the stage for our formalism, let us for definiteness consider the scattering process: $S_1(p_1) \ \chi_1(p_2)  \to S_2(k_1) \  \chi_2(k_2)$. At the tree level, this $2\to 2$ process may be represented by the $s$-channel Feynman diagram shown in Figure~\ref{fig:local_FeynD}. For later convenience, we define the total four-momenta, $p = p_1 + p_2$ and $k = k_1  + k_2$, of all particles in the initial and final state, respectively.
Applying the LSZ formalism~\cite{Lehmann:1954rq}, the transition amplitude $T_{\rm F}$, for the aforementioned process, may be evaluated as
\begin{align}
	T_{\rm F}\ =&\, -\lambda \, g \,\dint \dfrac{d^4 q}{(2\pi)^4} \ \dfrac{1}{q^2 - m_{\Phi}^2 + i \epsilon}
	\dint d^4 x \ d^4 y \  e^{-i (p - q ) \cdot x } e^{i (k - q) \cdot y}\nonumber\\
	=&\ \ \lambda \, g \, 
    \dfrac{\lrb{2 \pi}^4}{\absp^2 - \qt^2 - i \epsilon} \ \delta^{(4)}(p-k) \;,
	\label{eq:amplitude_std}
\end{align}
with $\qt^2 = (p^0)^2 - m_{\Phi}^2$. Note that energy-momentum conservation, $p = k$, arises as a result of Lorentz invariance and the local nature of the interactions. If the particle $\Phi$ obeys the on-mass shell (OS) condition in the $s$-channel of Figure~\ref{fig:local_FeynD}, we may naively incorporate its
decay width $\Gamma_\Phi$ by complexifying its squared mass~(e.g., see~\cite{Nowakowski:1993iu,Campagne:1997fu}), which amounts to making the substitution, $m^2_\Phi \to m^2_\Phi + i m_\Phi\Gamma_\Phi$, in~\eqref{eq:amplitude_std}.
For simplicity, we ignore possible finite width effects in this work, by setting $\Gamma_\Phi = 0$.

We should emphasise here that the transition amplitude $T_{\rm F}$ is not only Lorentz invariant, but also enjoys the fundamental property of analyticity which in turn implies the so-called crossing symmetry~\cite{Eden:1966dnq}. Specifically, the transition amplitude for the ($t$-channel) process $S_1(p_1) \  S_2(p_2) \to \chi_1(p^{\prime}_1) \  \chi_2(p^{\prime}_2)$ can be recovered from the $s$-channel amplitude given in~\eqref{eq:amplitude_std}. In this case, the relevant four-momenta, $p$ and $k$, are defined as: $p = p_1 - p_1^{\prime}$ and $k = p_2^{\prime} - p_2$. As a consequence of the analyticity of the S-matrix, we can only change the signs of the momenta of the incoming and outgoing particles, but the analytic structure of the amplitude in~\eqref{eq:amplitude_std} remains intact. This property of analyticity is that we wish to preserve in our formulation of a localised modification of the S-matrix which we discuss below.

\subsection{Analytic Localised Extension of the S-Matrix}

As stated earlier, the amplitude $T_{\rm F}$ in~\eqref{eq:amplitude_std}, as derived from the usual S-matrix theory, pertains to a scattering of particles with definite four-momenta. Hence, by virtue of the uncertainty principle${}$, no information about its space-time dependence is available. 
However, both the particles themselves and their interactions may be localised in a finite space-time volume. Following${}$~\cite{Kiers:1997pe,Ioannisian:1998ch}, we will assume the latter and regard all particles in the initial and final state of a process as being well described asymptotically by plane waves to a very good approximation. Such a consideration will be equivalent to the more often discussed wave-packet approach~(e.g., see~\cite{Akhmedov:2010ua}), since the localised interactions may be viewed as intersections of the wave packets of the initial and final particles at the vertices of a scattering process.

Let us first consider the \emph{production vertex} at some generic space-time point~$x$. To~introduce a finite non-zero  spread at $x$, we define the Lorentz-invariant Gaussian function~\cite{Ioannisian:1998ch,Naumov:2020yyv}
\begin{equation}
	G(x;\svev{x},\Delta p) = e^{ -\lrb{x-\svev{x}}^{\mu} \Delta p_{\mu \nu} \lrb{x-\svev{x}}^{\nu} } \;,
	\label{eq:general_smearing_def}	
\end{equation}
where the energy-momentum uncertainty tensor, $\Delta p_{\mu \nu}$,  is defined through the relation${}$: $\Delta p^{\rho\mu}\Delta x_{\mu\sigma} = \delta^{\rho}_{\sigma}$, with
\begin{equation}
	\Delta x_{\mu \nu} = \svev{x_{\mu}x_{\nu}} - \svev{x_{\mu}} \svev{x_{\nu}} \;.
	\label{eq:correlation_tensor}	
\end{equation}
Here, the parameters $\svev{x^{\mu}}$ and $\svev{x^{\mu}x^{\nu}}$ characterize the uncertainties in the four-position $x$. As we will see below, such finite uncertainties will trigger a nominal violation in the conservation of the four-momentum. However, this apparent violation should be treated with caution, and be interpreted instead as a non-conservation of the mean total momenta of the particles taking part in a localised scattering process. In fact, their momentum uncertainties, say $\delta p^\mu$, imply that the particles are not momentum eigenstates, so only their momentum mean values $p^\mu$ will be of physical relevance in our formalism.  Since the exponent of \eqref{eq:general_smearing_def} describes a complicated four-dimensional ellipsoid, we may simplify the analysis by assuming the factorisable form: $\Delta p^{\mu\nu} = \delta p^\mu \delta p^\nu$. In this case, we have   
\begin{equation}
	G(x;\xbL,\delta p) = e^{- \lrsb{(x-\xbL)\cdot\delta p}^{2}  } \;,
	\label{eq:spherical_smearing_x_def}	
\end{equation}
where $\xbL$ is the centre of the production vertex or the source, and $\delta p$ is the would-be four-momentum uncertainty. We note that $\delta p$ may naively be associated with an effective interaction radius $\delta x$ as $\dx^\mu = 1/\delta p^\mu$. 

By analogy, we may introduce the following localisation function for the \emph{detection vertex} at a generic four-position~$y$: 
\begin{equation}
	G (y;\ybL,\delta k) = e^{- \lrsb{(y-\ybL)\cdot\delta k}^{2}  } \;,
	\label{eq:spherical_smearing_y_def}	
\end{equation}
where $\ybL$ and $\delta k$ are the centre of the detection vertex and its would-be four-momentum uncertainty, respectively, with the effective interaction radius $\delta y$ defined as $\dy^\mu = 1/\delta k^\mu$. 

Taking into account the localisation functions in~\eqref{eq:spherical_smearing_x_def} and~\eqref{eq:spherical_smearing_y_def} for the four-positions $x$ and~$y$, the localised amplitude $T_{\rm L}$ for the process ${S_1\chi_1\to \Phi^*\to S_2\chi_2}$ takes on the form
\begin{align}
	T_{\rm L}(p,k; \xbL,\ybL,\delta x,\delta y)\, =&\ - \lambda \, g \dint d^4 x \, d^4 y \ e^{- \lrsb{(x-\xbL)\cdot \delta p}^{2} } \,  e^{- \lrsb{(y-\ybL)\cdot \delta k}^{2} }  \nonumber \\
	&\ \times e^{-i p \cdot x + i k \cdot y} \dint \dfrac{d^4 q}{(2\pi)^4} \ \dfrac{e^{i q \cdot (x-y)}}{q^2 - m_{\Phi}^2 + i \epsilon} \;,
	\label{eq:amplitude_init}
\end{align}
which, up to a constant, coincides with~\cite{Ioannisian:1998ch}. To be specific, the amplitude $T_{\rm L}$ describes the annihilation of the particles $S_1$ and $\chi_1$ with a sum of four-momenta $p$, at a mean four-position~$\langle x\rangle$ with an uncertainty~$\delta x$, and the subsequent creation of the particles $S_2$ and $\chi_2$ with a sum of four-momenta $k$, at a mean four-position~$\langle y\rangle$ with an uncertainty $\delta y$. In fact, such a setting may be applied equally well to describe particles that are forced to go through a restricted area of an aperture with a shape that is expressed by a function with a given localisation profile, e.g.~of the Gaussian form like in~\eqref{eq:spherical_smearing_x_def}. Hence, the localised S-matrix that we have been developing here can be viewed as another equivalent approach that allow us to describe particle diffraction, such as light diffraction, within the framework of QFT.        

\begin{figure}[t!]
	\centering\includegraphics[width=0.8\textwidth ]{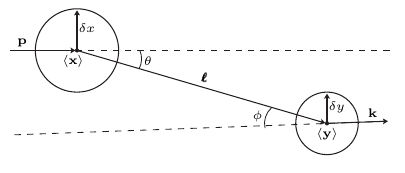}
	\caption{A schematic representation of the scattering process ${S_1\chi_1\to \Phi^*\to S_2\chi_2}$ with localised interactions, which corresponds to the transition amplitude $T_{\rm L}(p,k;\vecl ,\dx,\dy)$ in~\eqref{eq:amplitude_shift}.}
	\label{fig:non-local_FeynD}
\end{figure}

In various experiments, the system under study can be considered stationary to a good approximation. For example, in collider experiments, stochastic cooling~\cite{vanderMeer:312939} produces mono\-chromatic beams with small cross-sectional area. Confining particles via an external field means that the corresponding momentum operator does not commute with the interacting Hamiltonian of the system. This means that although the particles are approximate energy eigenstates, they suffer from three-momentum uncertainties. This is also the case in diffraction, where a monochromatic beam passes through an aperture of finite size. Similarly, in neutrino (or meson) oscillation experiments, time-uncertainties are much larger~\cite{Lipkin:1995cb,Grossman:1996eh} compared to spatial ones. Thus, any temporal uncertainty has to be included incoherently, at the amplitude-squared level.  
Therefore, for several applications of interest that we mentioned in the introduction, we may simplify our computations by adopting the working hypothesis that time uncertainties are much bigger than the corresponding spatial uncertainties, $\delta {\bf x}$ and $\delta {\bf y}$, and so take the infinite limit: $\delta x^0,\,\delta y^0 \to \infty$, or equivalently the zero limit: $\delta p^0, \, \delta k^0 \to 0$. This simplification will result in a localised transition amplitude~$T_{\rm L}$ which will be proportional to an irrelevant overall time-dependent phase. 
This stationary\- assumption, as well as the imposed spherical symmetry, render our model solvable, and so they allow our considerations to serve as a proof-of-concept of the LSMT. In general, beyond these assumptions, the amplitude can be computed using numerical methods.

As illustrated in Figure~\ref{fig:non-local_FeynD}, a further simplification occurs if all spatial uncertainties are taken to be equal, i.e.~$\delta x^i = \dx$ and  $\delta y^i = \dy$, for all $i=1,2,3$. Then, up to an overall frame-dependent phase factor $e^{i (\vecp \cdot \xb -  \veck \cdot \yb)}$, the localised amplitude for the process ${S_1\chi_1\to \Phi^*\to S_2\chi_2}$ becomes
\begin{align}
		T_{\rm L}(p,k;\vecl ,\dx,\dy)\, = &\ -2\pi\, \delta(p^0 - k^0) \ \lambda \, g \dint d^3 \bvec x \, d^3 \bvec y \ e^{-\bvec{x}^2/\dx^2 } \,  e^{-\bvec{y}^2/\dy^2 }  \nonumber \\
		&\ \times e^{i (\vecp \cdot \bvec x -  \veck \cdot \bvec y)} \dint \dfrac{d^3 \bvec q}{(2\pi)^3} \ \dfrac{e^{-i \bvec q \cdot (\bvec x- \bvec y - \vecl)}}{-|\bvec{q}|^2 + \qt^2  + i \epsilon} \;,
		\label{eq:amplitude_shift}
\end{align}
where we have introduced the average distance vector $\vecl \equiv \yb - \xb$. Notice that in addition to the momentum dependence of standard amplitude $T_{\rm F}$, the localised amplitude $T_{\rm L}$ now depends on the distance vector $\vecl$ between the production and detection vertices, and their uncertainties, $\dx,\, \dy$. Hence, the amplitude $T_{\rm L}$ is frame-independent as well, as a consequence of the Poincare invariance of the theory. Moreover, it is not difficult to see that in the infinite limits $\dx, \dy \to \infty $, the amplitude $T_{\rm L}$ in~\eqref{eq:amplitude_shift} becomes, up to an overall phase factor $e^{i{\vecp\cdot \vecl}}$, identical to the ordinary S-matrix amplitude $T_{\rm F}$ in~\eqref{eq:amplitude_std}.

As shown in~\cite{Ioannisian:1998ch} and in Appendices~\ref{app:amplitude},~\ref{app:angular} and~\ref{app:radial} using a different method, 
the various integrations over angular and radial variables in~\eqref{eq:amplitude_shift} can be performed analytically, yielding the amplitude
\begin{align}
T_{\rm L}(p,k; \vecl,\dx,\dy)\, =&\ 2\pi\, \delta(p^0 - k^0) \ \lambda\, g\, \dfrac{\pi^2}{8}\,
\dfrac{ \dx^3 \,  \dy^3}{\absL}\;  
	e^{-\lrsb{\lrb{|\vecp|^2+\qt^2} \dx^2 +\lrb{|\veck|^2+\qt^2} \dy^2 }/4} \nonumber \\  
	&\ \times \bigg( e^{i \qt\,\absL} \, \Erfc\,z_-\: -\: e^{-i \qt\,\absL} \, \Erfc\,z_+ \bigg)\;.
	\label{eq:amplitude_final}
\end{align}
In the above, we have used the shorthand notation: $\vecL = \vecl- \frac{i}{2} \big(\vecp\,\dx^2 + \veck\, \dy^2\big)$, $\absL \equiv \sqrt{\vecL \cdot \vecL}$, and $z_\pm = - \dfrac{i}{2} \qt \, \dl\, \pm\, \dfrac{\absL}{\dl}$, with $\dl^2 = \dx^2 + \dy^2$. In~\eqref{eq:amplitude_final}, $\Erfc\,z$ is the complementary error function analytically continued with a complex argument $z \in \mathbb{C}$ as follows:
\begin{equation}
    \label{eq:Erfc}
    \Erfc\,z \: =\: 1\, -\, \frac{2}{\sqrt{\pi}}\, 
    \int_0^z\,dt\, e^{-t^2}\, . 
\end{equation}

We should point out that the localised amplitude $T_{\rm L}$ for the process $S_1\chi_1\to \Phi^*\to S_2\chi_2$ remains finite in the OS kinematic region, $|\vecp| = |\veck| = \qt$, as long as $\dl$ is finite, even if one assumes a vanishing $\Phi$-decay width $\Gamma_\Phi$. On the other hand, due to the analyticity of~$T_{\rm L}$, the same analytic expression in~\eqref{eq:amplitude_final} may be used to regulate the $t$-channel singularities~\cite{Coleman:1965xm,Peierls:1961zz,Ginzburg:1995bc},
which can occur in the crossing symmetric process ${S_1S_2 \to \Phi^* \to \chi_1\chi_2}$ in the physical region. Unlike other methods that model the finite size of the interacting beams~\cite{Kotkin:1992bj,Melnikov:1996na,Melnikov:1996iu}, LSMT takes into account the finite size of the interaction volume coherently, where all spatial uncertainties are implemented at the amplitude level~$T_{\rm L}$, and not at its square~$|T_{\rm L}|^2$. As well as being devoid of $t$-channel infinities, the amplitude $T_{\rm L}$ also contains information for the
distance between the source and the detector, through the distance vector $\vecl$. The latter can shed light on phenomena that may take place on both microscopic and macroscopic distances, like neutrino oscillations~\cite{Ioannisian:1998ch}, which we discuss in more detail in Sections~\ref{sec:approx} and~\ref{sec:numerics}.

We note that the localised amplitude given in~\refs{eq:amplitude_final} is finite at $\absL = 0$. This can be easily deduced by observing that $T_{\rm L}$ is an even function with respect to $\absL$, implying that $\absL\ T_{\rm L}$ is an odd one. Then, by means of a Taylor series expansion, one may verify that $\absL\ T_{\rm L}$ approaches zero at least as fast as $\absL$, so $T_{\rm L}$ is finite at $\absL = 0$.

As shown in ref.~\cite{Ioannisian:1998ch}, the computation of the amplitude for fermions is analogous to the scalar case we study here. Therefore, if we consider particles at different Lorentz representations, the corresponding amplitude has a similar form to~\refs{eq:amplitude_final} with some extra factors that depend on the Lorentz structure of the particles. Thus, our conclusions can be extended to other cases.
    
Although the process we study is simple with only one Feynmann diagram, the results we obtain are fairly general. This formalism can explain spatial oscillations of mixed mediators~\cite{Ioannisian:1998ch}, regularise $t$-channel singularities, and shows a correspondence between QFT and diffraction. Introducing more particles, interactions, or channels increases the complexity of the computation, but we do not expect that complexity will spoil the analytic features of the formalism under study.

We conclude this section by presenting two more interesting limits concerning $T_{\rm L}$ stated in~\refs{eq:amplitude_final}.

\subsubsection{Zero-spread  limit}

In the limit of vanishing spread of the production and detection vertices, \ie ${\dx, \dy \to 0}$, the complementary error functions take the values: $\Erfc\, z_+ \to 0$ and $\Erfc\,z_- \to 2$. In this vanishing limit of $\dl$, the localised amplitude~\refs{eq:amplitude_final} simplifies to
\begin{equation}
	T_{\rm L}(p,k; \vecl)\, =\,  2\pi \, \delta(p^0 - k^0) \ \lambda \, g\; \dfrac{\pi^2}{4} \dx^3 \,  \dy^3 \ \dfrac{  e^{i \qt \absl}}{\absl}\;.
	\label{eq:amplitude_zero_volume}
\end{equation}
This form of the amplitude implies that the exchanged particle passes through definite points in space, $\xb$ and $\yb$. In fact, 
$T_{\rm L}$ gets proportional to the Green's function of the Euclidean 3D space, $e^{i \qt \absl}/\absl$. Since the momentum 
uncertainties diverge when $\dl\to 0$, the resulting incoming and outgoing momenta, ${\bf p}$ and ${\bf k}$, will be unrelated to each other and so arbitrary. However, the three-momentum of the mediator may be identified by its wavenumber, $\qt$. If  $\qt^2 \geq 0$, this would correspond to a real particle with momentum $\qt$. Instead, for $\qt^2 < 0$, the amplitude $T_{\rm L}$ would fall off exponentially as $e^{-|\qt|\, \absl}$, representing a decaying mode that travels an effective mean distance of~$1/|\qt|$ from a point source.

\subsubsection{Momentum conservation limit}

For most experimental settings, we expect $\absp \, \dl , \; \absk \, \dl \gg 1$, so that the violation of energy-momentum conservation is marginal, with $\vecp \simeq \veck$. In principle, we may enforce a total four-momentum conservation limit by assuming a translationally invariant localisation of the form 
$e^{-\lrb{\bvec{x}-\bvec{y}-\vecl}^2/\dl^2}$,
instead of two independent Gaussians centered at $\langle {\bf x}\rangle$ and $\langle {\bf y}\rangle$.
Upon integration over the coordinates, the above restricted form of the smearing profile gives rise to the 3D $\delta$-function, $\delta^{(3)}\lrb{\vecp -\veck}$, in~\refs{eq:amplitude_final}.  This ensures four-momentum conservation between the incoming
and outgoing particles, i.e.~$\vecp = \veck$, even though one still has in general ${\bf q} \neq \vecp$  for the momentum ${\bf q}$ of the exchanged particle $\Phi$. In addition, the  overall constant changes by a factor $\dfrac{1}{\pi^{3/2}}\dfrac{\dl^3}{\dx^3 \dy^3}$.

Putting everything together, the localised amplitude $T_{\rm L}$ reads 
\begin{align}
	T_{\rm L}(p,k; \vecl,\dl)\, =&\ (2\pi)^4 \, \delta(p^0 - k^0)\ \delta^{(3)}(\vecp - \veck)\  \lambda \, g\ \dfrac{\sqrt{\pi}}{8}\, \dfrac{ \dl^3 }{\absL}\;  
	e^{-\dl^2 \lrb{|\vecp|^2+\qt^2}/4   } \nonumber \\  
	&\ \times \bigg( e^{i \qt\, \absL} \, \Erfc\,z_-\: -\: e^{-i \qt\, \absL} \, \Erfc\,z_+ \bigg)\;,
	\label{eq:amplitude_k=p_final}
\end{align}
where $\vecL = \vecl- \frac{i}{2} \,\vecp\, \dl^2$, and $\dl$ and the complex arguments $z_\pm$ are defined after~\eqref{eq:amplitude_final}. Without compromising the main features of our localised S-matrix formalism, we shall employ the
simplified  amplitude $T_{\rm L}$ given by~\eqref{eq:amplitude_k=p_final}. To further simplify matters, we strip off an overall factor of $(2\pi)^4\, \delta^{(4)}(p - k)\, \lambda\,g$ from $T_{\rm L}$ in~\eqref{eq:amplitude_k=p_final},
and define a corresponding localised matrix element~$M$ as follows:
\begin{equation}
	M(\vecp,\qt; \vecl,\dl)\ =\ \dfrac{\sqrt{\pi}}{8}\, \dfrac{ \dl^3 }{\absL}\;  
	e^{-\dl^2 \lrb{|\vecp|^2+\qt^2}/4}\,
	\bigg( e^{i \qt\, \absL} \, \Erfc\,z_-\: -\: e^{-i \qt\, \absL} \, \Erfc\,z_+ \bigg)\;.
	\label{eq:matrix_element_def}
\end{equation}
Our analysis in the following two sections will utilise this last form of the matrix element $M$.

\section{Near- and Far-Field Approximations}\label{sec:approx}
\setcounter{equation}{0}

Although the matrix element $M$ in~\refs{eq:matrix_element_def} for a generic localised process $S_1 \chi_1\to \Phi^*\to S_2 \chi_2$ is given in a closed form, it still remains difficult to deduce from the latter what its main physical implications are. To better understand these, we derive in this section analytical approximations of $M$ as a function of the distance vector $\vecl$ between the production of the $\Phi$-mediator and its detection, the spatial distance uncertainty $\dl$, as well as of the total three-momentum ${\bf p}$ of the incoming particles $S_1$ and $\chi_1$. In all our approximations, we consider that $\absp \, \dl \gg 1$, which happens to be a valid assumption for most realistic situations. In close analogy with diffractive optics, we differentiate two regions: (i)~the Fraunhofer or far-field zone where $\absl \gg  \absp \, \dl^2$, and (ii)~the Fresnel or near-field zone in~which $\absl \ll \absp \, \dl^2$. 

Depending on the magnitude of $z_{\pm}$ of the complementary error function $\Erfc\,z$ defined in~\eqref{eq:Erfc}, we may use either a Taylor series expansion~\cite{HandbookCF:2008qt},
\begin{equation}
\Erfc \, z\, \simeq\, 1- \dfrac{2}{\sqrt{\pi}} z  \,,  \label{eq:erfc_taylor}
\end{equation}
for $|z| \ll 1$, or an asymptotic expansion~\cite{HandbookCF:2008qt},
\begin{equation}
\Erfc\,z\, \simeq\ \dfrac{e^{-z^2}}{\sqrt{\pi} \ z} \ ,
	\label{eq:erfc_asymptotic}
\end{equation}
when $|z| \gg 1$ and $\arg z < 3\pi/4$. If $\arg z \geq 3 \pi/4$, the asymptotic expansion may be obtained after applying first the identity: $\Erfc\,z = 2 - \Erfc(-z)$.

\subsection{Fraunhofer Zone}

In the Fraunhofer or far-field region, the distance is $\absl \gg \absp \, \dl^2$, so the complex vector norm $\absL$ may then be approximated as
\begin{equation}
	\absL\, \simeq\, \absl - \dfrac{i}{2} \cos\theta \ \absp \ \dl^2\, -\, 
 \dfrac{\sin^2\theta \ \absp^2\, \dl^4}{8\,\absl} \, , 
	\label{eq:L_far}
\end{equation}
where $\theta$ is the angle between $\vecl$ and $\vecp$. In this limit, we can ignore the exponentially suppressed term of~\refs{eq:erfc_asymptotic}. Thus, $\Erfc\,z_- \simeq 2$ and $\Erfc\,z_+ \simeq 0$, and the matrix element reads
\begin{equation}
	M\, \simeq\ \dfrac{\sqrt \pi}{4} \; \dl^3 \ \dfrac{e^{i \qt \absl}}{\absl}\;  e^{-\lrb{\vecp - \qt \; 
 \hatl}^2\dl^2/4}  \;,
	\label{eq:M_far}
\end{equation}
with $\hatl$ being a unit vector along the distance vector $\vecl$. The square of the matrix element $M$ in~\eqref{eq:M_far} obeys the expected inverse-square law $1/|\vecl|^2$ for $\qt^2 \geq 0$, \ie~for physical intermediate states. This result is in agreement with the so-called Grimus-Stockinger theorem~\cite{Grimus:1996av}, which is only applicable in the Fraunhofer regime~\cite{Ioannisian:1998ch,Akhmedov:2009rb,Akhmedov:2010ms,Naumov:2013bea}. 

For off-shell particle virtualities with $\qt^2 < 0$, the approximate matrix element $M$ in~\eqref{eq:M_far} can be analytically continued from $\qt \to i|\qt|$, and so one can show that $M$ falls off exponentially, i.e.~$M \propto e^{- |\qt| \absl} / \absl$. We~note that this exponential fall-off of $M$ with increasing distance $\absl$ is much stronger than the generic weaker scaling behaviour of $M \propto \absl^{-2}$, claimed in~\cite{Grimus:1996av,Grimus:2019hlq}. Furthermore, it is not difficult to see from~\eqref{eq:M_far} that for $\qt^2 <0$, one has $M \propto \exp\big[\,i|\qt|\, \vecp \cdot \hatl\, \dl^2/2\big]$, but this only surviving phase leads to no favourable direction on the particle propagation in the Fraunhofer zone, i.e.~$|M|$ is completely independent of~$\theta$. 

We should also observe that for finite spatial uncertainties $\dl$, the  localised matrix element${}$~$M$ is devoid of the $s$-channel singularity${}$ haunting the ordinary S-matrix amplitude $T_{\rm F}$ in~\eqref{eq:amplitude_std}, in the OS limit $\absp \to \qt$. Further more, for a given angle $\theta_*$, there is a characteristic momentum, which we call~$\pmax$, that maximizes the norm of~$M$, $|M|$. In particular, we find that $\pmax$ is shifted from its OS value $\pmax = \qt$ in the forward direction  to smaller values, according to the simple relation: $\pmax = \qt \, \cos\theta_*$. Such shifts may be probed in observations that would involve non-zero angles $\theta$, and as such, they may provide a non-trivial test of LSMT under study.

Finally, it is interesting to remark that if the $\Phi$-mediator has real momentum ($\qt^2>0$), the localised amplitude $M$ will be suppressed away from the forward ($\theta=0$) direction, because of the exponential factor $e^{-\lrb{\vecp - \qt \; \hatl}^2\dl^2/4}$ in~\eqref{eq:M_far}. This factor also disfavours  propagation in the backwards hemisphere for angles~$\theta \ge \pi/2$, and so it resembles the engineered \emph{obliquity factor} that features in the well-known Helmholtz--Kirchhoff diffraction formula (see, \eg~\cite{Optics_1999}). 
But~unlike the classical case, the LSMT provides naturally the necessary ``quantum obliquity factor'' which although it suppresses, it does not prohibit particle propagation in a classically forbidden region. As we will see in the next subsection, this property still holds true for the Fresnel region as well.

\subsection{Fresnel Zone}
\begin{table}[t]
    \centering
    \begin{tabular}{|c||l|c|}
    \hline
        Subregion & \hspace{25mm}Conditions & 
        Magnitude of $z_-$ \\
        \hline\hline
       \textbf{I} & $\big|\absp -\qt\big|\,\dl \gg {\rm max}\big(1,\;
        |\hatp\cdot\vecl|/\dl \big)$                 & $|z_-| \gg 1$ \\ \hline
       \textbf{II} & 
        $|\hatp\cdot\vecl| \ll \dl$ ~and~ $\big|\absp -\qt\big|\,\dl \ll 1$                    & $|z_-| \ll 1$ \\ \hline
       \textbf{III} & $|\hatp\cdot\vecl| \gg \dl$ ~and~ $\big|\absp-\qt\big|\,\dl^2 \ll |\hatp\cdot\vecl|$ & $|z_-| \gg 1$ \\ \hline 
    \end{tabular}
    \caption{The three Fresnel subregions as described in more detail in the text. The conditions which hold in all subregions are: $\absp \, \dl^2 \gg \absl$ and $\absp \,\dl \gg 1$. Note that the latter entails $|\absp + \qt |\dl \gg 1$, which in turn implies $|z_+| \gg 1$ for all subregions.}
    \label{tab:fresnel_Subregion}
\end{table}

In the Fresnel or near-field region, in which $\absl \ll \absp\, \dl^2$,  the complex norm $\absL$ may be expanded as
\begin{equation}
	\absL\: \simeq\: -\, \dfrac{i}{2} \absp \, \dl^2\, +\,
 \absl  \, \cos\theta\, +\, \dfrac{i\,\absl^2}{\absp \, \dl^2} \, \sin^2\theta\, , 
		\label{eq:L_near}
	\end{equation}
when $\cos\theta <0$. Although $\absL$ should be multiplied by $-1$ for $\cos\theta \geq 0$, we can still use \refs{eq:L_near}, since the amplitude~\refs{eq:amplitude_final} is an even function of $\absL$.

Given the central working hypothesis $\absp \, \dl \gg 1$, it follows that $\big|\absp+\qt\big|\dl \gg 1$. This in turn implies that $|z_+| \gg 1$. Consequently, $\Erfc \,z_+$ can be expanded as in~\refs{eq:erfc_asymptotic}. On the other hand, the size of $|z_-|$ depends on the magnitude of the dimensionless quantities: $|\hatp\cdot\vecl|/\dl$ and $\big|\absp-\qt\big|\,\dl$, where $\hatp \equiv \vecp/\absp$ is a unit vector along the three-momentum $\vecp$. The first quantity, $|\hatp\cdot\vecl|/\dl$, gives a measure of the projection of $\vecl$ onto the direction of $\vecp$ in units of $\dl$. The second one, $\big|\absp-\qt\big|\,\dl$, quantifies the degree of ``off-shellness'' of the exchanged $\Phi$ particle.

The various possible hierarchies of the two quantities, $|\hatp\cdot\vecl|/\dl$ and $\big|\absp-\qt\big|\,\dl$, 
form three subregions. These are succinctly summarized in Table~\ref{tab:fresnel_Subregion}. 
In detail, Subregion~\textbf{I} is defined by the constraint: $\big|\absp -\qt\big|\,\dl \gg {\rm max}\lrb{ 1,\;\big|\hatp\cdot\vecl\big|/\dl }$, which implies that $|z_-| \gg 1$. 
Subregion~\textbf{II} corresponds to $|\hatp\cdot\vecl| \ll \dl$ and $\big|\absp-\qt\big|\dl \ll 1$, and so it is $|z_-| \ll 1$. 
Finally, Subregion~{\bf III} is given by $\big|\hatp\cdot\vecl\big| \gg \dl$ and $\big|\absp-\qt\big|\dl^2 \ll |\hatp\cdot\vecl|$, which results in $|z_-| \gg 1$. 
Notice that $\big|\absp-\qt\big|\,\dl \ll 1$ defines a resonant region for the $\Phi$ mediator. But as happened in the Fraunhofer zone, the maximum of the modulus of the matrix element, $|M|$, is not guaranteed to occur on the resonance point, $\absp=\qt$, as the angle $\theta$ varies from 0 to $\pi$.

\subsubsection{Subregion~\textbf{I}: 
\texorpdfstring{$\big|\absp -\qt\big|\,\dl \gg {\rm max}\big(1,\;|\hatp\cdot\vecl|/\dl\big)$}{subI} 
}
In this subregion, $\Erfc \, z_\pm$ can be approximated as in~\refs{eq:erfc_asymptotic}. Then, the matrix element becomes
\begin{align}
	M\, \simeq&\ \: \dfrac{\dl^3}{8\,\absL} \lrb{\dfrac{1}{z_-}-\dfrac{1}{z_+}} e^{i  \vecp \cdot \vecl\: -\: \lrb{\absl/\dl}^2 }\, \simeq\, 
	\dfrac{\dl^2 \ e^{i  \vecp \cdot \vecl - \lrb{\absl/\dl}^2 }}{\Big(\absp^2 - \qt^2\Big) \dl^2\: 
 +\: 4 \Big[ i \vecp \cdot \vecl - \big(\absl/\dl\big)^2\Big]}\ .  
	\label{eq:M_sub-I} 
\end{align}
We should note that this approximation is accurate up to corrections $\mathcal{O}\big[\dl^2\,|\vecp \times \vecl|^4/(\absp \, \dl)^6 \big]$. There are other relevant higher order terms that depend on the sign of $\qt^2$ and $\vecp \cdot \vecl$, as well as on their relative size. For example, when both $\hatp\cdot \vecl$ and $\qt^2$ are negative with $|\qt| \dl^2/2 < |\hatp\cdot \vecl|$,  higher order terms $\mathcal{O}\Big(\exp\big[ |\qt| \, \hatp\cdot \vecl + \qt^2\dl^2/2\big] \Big)$ that may potentially appear are getting suppressed by their negative exponent.
Finally, like in the Fraunhofer region for $\qt^2 < 0$, there is no directional constraint on $|M|$ in this subregion.

In Subregion~{\bf I}, for $\qt \, \dl^2 \gg \absl$, the characteristic momentum $\pmax$ that maximizes $|M|$ obeys the relation: $\pmax \simeq \qt + 2(1-2\cos^2\theta)\,\absl^2/(\qt\;\dl^4)$. Thus, we have $\pmax > \qt$ ($\pmax < \qt$) for $|\cos\theta|<1/\sqrt{2}$ ($|\cos\theta|>1/\sqrt{2}$). On the other hand, if $\qt \, \dl^2 \ll \absl$, the momentum that maximizes the matrix element turns out to be: $\pmax \simeq \sqrt{|\cos 2\theta|}\,\absl/\dl^2$, which does not respect Fresnel's central constraint: $\absp\,\dl^2 \gg \absl$. In this case, the matrix element in Sub-region~{\bf I} does not exhibit a maximum. Instead, it decreases monotonically with the momentum, i.e.~${|M|\propto 1/\absp^2}$. We note that for $\cos\theta=0$, the matrix element $M$ as approximated in~\refs{eq:M_sub-I} appears to have singularities, when $\absp^2 =\qt^2 +\, 4\absl^2/\dl^4$. However, these would-be singularities are not present. They originate from $z_{\pm} =0$, and so they violate  the basic assumption $|z_\pm| \gg 1$ that underlies the validity of this approximation.

\subsubsection{Subregion~\textbf{II}: 
\texorpdfstring{$|\hatp\cdot\vecl| \ll \dl$ and $\big|\absp -\qt\big|\,\dl \ll 1$}{subII}
} 
If the momenta obey the resonant condition,
$\big|\absp -\qt\big|\,\dl \ll 1$, and also $|\hatp\cdot\vecl| \ll \dl$, we then have $|z_-|\ll 1$. Making use of~\eqref{eq:erfc_taylor}, the matrix element may be approximated as
\begin{align}
    M\, \simeq&\ \,
    \dfrac{ i \sqrt \pi}{4}  \dfrac{\dl}{\absp}\:
	\bigg[1+\frac{2}{\sqrt{\pi}} \ \dfrac{ \hatp\cdot \vecl  }{\dl} - \dfrac{i}{\sqrt{\pi}} \big(\absp - \qt\big)\dl\bigg] \nonumber \\[2mm] 
	&\ \times \exp\bigg[\!-\frac{1}{4}\big(\absp-\qt\big)^2\dl^2\, +\, \qt\,\bigg( i \, \hatp\cdot \vecl - \frac{|\hatp \times \vecl|^2 }{\absp\,\dl^2}\,\bigg)\bigg]  
	\nonumber \\[2mm]
	&\ +\: \dfrac{1}{2}\, \dfrac{e^{i  \vecp \cdot \vecl-(\absl/\dl)^2}}{ \absp \, \big(\absp+\qt\big)}
    \;.
    \label{eq:M_sub-II} 
\end{align}
We observe that for a finite $\dl$, the singularity of the S-matrix amplitude $T_{\rm F}$ in~\eqref{eq:amplitude_std} 
at $\absp = \qt$ is successfully regulated. Furthermore, the value of the matrix element $M$ in the OS limit, $\absp \to \qt$, gets reduced as $\theta$ increases, with a minimum in the backwards direction $\theta =\pi$. On~the other hand, for $\theta \to 0$, $|M|$ increases slightly with the distance $\absl$. Thus, there seems to be a focusing effect that makes the observation (or decay) more probable away from the origin, $\absl=0$. Also, this phenomenon may affect the assumed flux for the $\Phi$ particles 
at the source. 

The value of the characteristic momentum $\pmax$ that gives rise to a maximum $|M|$ is estimated to be 
\begin{equation}
    \pmax\: \simeq\: \qt\, +\, \dfrac{2}{\qt\,\dl^2} \ \lrb{\dfrac{|\hatp \times \vecl|^2}{\dl^2} - 1}\;.
    \label{eq:pmax_sub-II}
\end{equation}
Although this estimate assumes $\big|\absp -\qt\big|\,\dl \ll 1$, the resulting value of $\pmax$ may lie outside or be at the boundary of this Fresnel subregion. In such case, $M$ should be estimated numerically using~\refs{eq:matrix_element_def}, as done in Section~\ref{sec:numerics}. However, the above exercise is still useful as it shows that the maximum occurs at $\pmax$ that may be below and above $\qt$, for $|\hatp \times \vecl| < \dl$ and $|\hatp \times \vecl| > \dl$, respectively. 

We must remark that the approximate matrix element in~\refs{eq:M_sub-II} offers a rather accurate description of the exact amplitude $M$ in Subregion~{\bf II}. The main higher order contribution is~$\mathcal{O}\big[|\hatp \cdot \vecl|/(\absp^2\,\dl)\big]$. All other higher order corrections turn out to be subdominant.

\subsubsection{Subregion~\textbf{III}: 
\texorpdfstring{$|\hatp\cdot\vecl| \gg \dl$ and $\big|\absp-\qt\big|\,\dl^2 \ll |\hatp\cdot\vecl|$}{subIII} 
}
If the detection vertex obeys the restrictions: $|\hatp\cdot\vecl| \gg \dl$ and $\big|\absp-\qt\big|\,\dl^2 \ll |\hatp\cdot\vecl|$, the complementary error functions are then expanded as in~\refs{eq:erfc_asymptotic}. Despite $|z_-| \gg 1$ in both Subregions~{\bf I} and~{\bf III}, we find that  the matrix element assumes different forms as different terms dominate in the expansion of the arguments $z_\pm$. Hence, in Subregion~{\bf III} the matrix element will be approximated as 
\begin{align}
	M\, \simeq\ \; & 
	\pm \dfrac{i\dl^2\, \sqrt{\pi}}{2}\; \dfrac{ \exp\bigg[\!-\dfrac{1}{4}\big(\absp \mp \qt\big)^2\, \dl^2\: \pm\: \qt\, \bigg(i\hatp\cdot\vecl\, -\, \dfrac{|\hatp\times\vecl|^2}{\absp\, \dl^2}\bigg)\bigg] } {\absp \; \dl\: +\: 2 i\,\hatp \cdot \vecl/\dl} \nonumber\\[2mm]
	& +\
	\dfrac{\dl^2 \, e^{i\vecp\cdot\vecl  -  \lrb{\absl/\dl}^2  }}{\big(\absp^2 - \qt^2\big)\,\dl^2 \:  +\: 4\lrsb{i\vecp\cdot\vecl\: -\: \big(\hatp\cdot\vecl/\dl\big)^2} }	\;.
	\label{eq:M_sub-III}
\end{align}
In the above, the upper (lower) sign corresponds to  $\cos\theta > 0$ ($\cos\theta < 0$), \ie~towards the forward (backward) direction, and originates from the second (first) term of~\refs{eq:matrix_element_def}. Instead, the last term 
in~\eqref{eq:M_sub-III} remains the same in both directions. Evidently, as $|\vecp \cdot \vecl| \gg \absp\,\dl \gg 1$, it~is not difficult to verify that the matrix element in~\eqref{eq:M_sub-III} is finite in the resonant region $\absp \simeq \qt$. 

In this subregion, the angular dependence is more involved than that in the other two. However, backwards particle propagation gets strongly disfavoured within Subregion~\textbf{III}. We may elucidate this by first  considering propagation in the forward direction, $\theta = 0$. In this case, the parameters satisfy: $\absl \gg \dl$, $\absl \ll \absp \, \dl^2$, and $\big|\absp -\qt\big|\,\dl^2 \ll \absl$. Under these conditions,~\refs{eq:M_sub-III} is dominated by its first term which is only slightly reduced as the distance increases. On the other hand, for $\theta = \pi$, the second term in~\eqref{eq:M_sub-II} will become dominant. In~this case, however, propagation in the backwards direction will be disfavoured, because of the exponential suppression factor $e^{-(\absl/\dl)^2}$. These attributes will be discussed in more detail in Section~\ref{sec:numerics}, where the exact matrix element~\refs{eq:matrix_element_def} will be numerically evaluated.   

In Subregion~{\bf III}, the characteristic momentum, $\pmax$, that maximizes (locally) $|M|$ depends on both the angle~$\theta$ and the average distance~$\absl$. To explicitly demonstrate this dependence, we consider again the forward and backward directions which have $\theta=0$ and $\theta=\pi$, respectively. In the former, we have $\pmax \simeq \qt -2/(\qt \, \dl^2)$, while it is $\pmax \simeq \qt - 2\absl^2/(\qt\,\dl^4)$ in the latter.   Although the shift of the maximum is negative, its magnitude in the backwards direction is enhanced by a factor of $(\absl/\dl)^2$. We note that this enhancement should not be fully trusted, as it originates from a $\pmax$ whose value lies outside or is on the boundary of this subregion, such that $\big|\pmax -\qt\big|\,\dl^2\geq |\hatp \cdot \vecl|_{\rm *}$.

The matrix element~\refs{eq:M_sub-III} is obtained by ignoring several higher order corrections. For example, the first term is accurate up to $\mathcal{O} \big[\absl/(\absp^2 \, \dl^3)\big]$. Although such terms are found to be subdominant, they need to be included in order to obtain an accurate numerical value for the matrix element${}$~$M$. Nevertheless, we find that the relative numerical difference by evaluating the two expressions in~\refs{eq:M_sub-III} and~\refs{eq:matrix_element_def} is typically within the $20\%$ level.  

\subsubsection{Standard S-matrix limit in the Fresnel region}

The standard S-matrix limit is a special case of Subregions~\textbf{I} and~\textbf{II}. For the former, taking the limit $\absl/\dl \to 0$ is a straightforward exercise. For the latter, thanks to the $\delta$-function representation
\begin{equation}
	\displaystyle \lim_{\epsilon \to \infty} \epsilon\ e^{ - \epsilon^2 \, x^2 }  \to  \sqrt{\pi} \delta(x) \;,
	\label{eq:delta_def}
\end{equation}
it can be shown that 
\begin{equation}
	M\, =\   \dfrac{ i \pi}{2}  \dfrac{ e^{i \qt\, \hatp\cdot \vecl }}{\absp} \ \delta(\absp - \qt)   	  \;.
	\label{eq:M_res_dl_inf_start}
\end{equation}
This last expression can be rewritten as
\begin{equation}
	M\, =\   i \pi \  e^{i \vecp \cdot \vecl} \ \delta_{+}(\absp^2 - \qt^2) \;,
	\label{eq:M_res_dl_inf_final}
\end{equation}
where $\delta_{+}(\absp^2 - \qt^2) \equiv \delta (\absp^2 - \qt^2)\: \theta (\absp )$. 

Finally, as $\dl / \absl \to \infty$ for any finite value of $\absl$, we may combine the approximate expressions in~\refs{eq:M_sub-I,eq:M_sub-II} in order to write the matrix element into the more familiar form (e.g., see~\cite{Schwartz:2014sze}):
\begin{equation}
	M\, =\, \    e^{i \vecp \cdot \vecl} \ \bigg[ i \pi \ \delta_{+}(\absp^2 - \qt^2) + \mathcal{P}\bigg\{\dfrac{1}{\absp^2 - \qt^2}\bigg\} \bigg] \;,
	\label{eq:M_dl_inf}
\end{equation}
where $\mathcal{P}\{ \, \dots\}$ denotes the Cauchy principal value. Also, notice that the appearance of an overall (unobservable) $\vecl$-dependent phase in~\eqref{eq:M_dl_inf} due to the spatial translation invariance of the original localised amplitude in~\refs{eq:amplitude_shift}. Otherwise, the matrix element~$M$ in~\eqref{eq:M_dl_inf} matches exactly with the standard result of the S-matrix amplitude in~\eqref{eq:amplitude_std}.

\subsubsection{Oscillations of mixed mediators in the Fresnel region}

The approximations,~\refs{eq:M_sub-II,eq:M_sub-III}, of the matrix element in the Fresnel Subregions~\textbf{II} and~\textbf{III} indicate  that the propagation of mixed mediators will also give rise to their oscillation in this regime, where ${\absp \, \dl^2 \gg \absl}$. First, we should observe that exponential suppression factors, such as those mentioned in~\cite{Beuthe:2001rc}, play no role here, as they are direction-dependent and vanish not only in the forward direction ($\theta = 0$), but also in the backward direction ($\theta = \pi$). 

Let us have a closer look at the phenomenon of oscillations within Subregion~\textbf{III} in the forward direction~(${\theta =0}$). As a mixed system of mediators, we may consider two exchanged particles, $\Phi_{1,2}$, with different masses, $m_{1,2}$, obeying the hierarchy: $0< \Delta m = m_2 - m_1 \ll m_{1,2}$. After setting all coupling constants of the theory to 1, for simplicity, the total matrix element $M$ will be the sum of two matrix elements, $M_{1,2}$, describing the exchange of the particles $\Phi_{1,2}$ in the $s$-channel, i.e.~$M = M_1 + M_2$. Each matrix element~$M_{1,2}$ can then be expanded according to~\refs{eq:M_sub-III}, with $\qt^2_{1,2} = (p^0)^2 - m^2_{1,2}$. Although the complex norms $|M_{1,2}|$ individually do not predominantly depend on $\absl$, there is still a phase difference between $M_1$ and $M_2$, given by $\exp\big[ i(\qt_1 - \qt_2)\,\hatp\cdot \vecl\big]$. Obviously, this phase difference induces an oscillating pattern in $|M|$ with oscillation length: ${L_\text{osc} = |\qt_1 - \qt_2|^{-1}}$. This pattern is exactly the same as in the frequently discussed Fraunhofer zone, but it has an almost constant amplitude $|M|$ with the distance $\absl$ like \emph{plane waves} as observed in~\cite{Ioannisian:1998ch,Naumov:2013bea}, rather than it is decreasing as $1/\absl$ as spherical waves. A~similar conclusion would be reached if we had considered Subregion~{\bf II}, which represents a region in the deep Fresnel zone, as it lies much closer to the QM center~$\xb$ of the source. Here, we must caution the reader that statistical uncertainties, $\sigma_{\vecl}$, play a significant role in oscillations. These are usually larger than~$\dl$, i.e.~$\sigma_{\vecl} \gtrsim \dl$, and so they will reduce the amplitude of oscillations, at least by a factor $L_{\rm osc}/\sigma_{\vecl} \ll 1$, in oscillation scenarios with $L_{\rm osc} \ll \dl$~\cite{Ioannisian:1998ch,Beuthe:2001rc}. Statistical uncertainties are a source of decoherence with $\sigma_{\vecl} \gtrsim \dl$, which means that the features we pointed out in the Fresnel region may be obscured. However, the amplitude will still be finite, because $\sigma_{\vecl}$ cannot introduce singularities. 

In the backwards direction ($\theta = \pi$) of Subregion~{\bf III}, the last term on the RHS of~\refs{eq:M_sub-III} will dominate, and so $|M|=|M_1+M_2|$ will have a tiny oscillating amplitude. As a result, there will be no visible oscillations in this region. As for Subregion~{\bf I}, it is worth commenting that it is a kinematic region signifying a highly off-shell regime of particle propagation, since we have the condition: $\big|\absp - \qt\big| \gg \absl/\dl^2$, specifically~in the forward (backwards) direction where~${\theta = 0\,(\pi)}$. According to the analytic matrix-element approximation in~\eqref{eq:M_sub-I}, no oscillations from mixed mediators will take place in this subregion.

In conclusion, the predictions derived from our LSMT can be tested against experiments designed to measure directional dependence of interactions, as well as particle oscillations.
The latter may not only take place in the usually considered Fraunhofer region which lies far away from the source, but also within the Fresnel zone as we have explicitly demonstrated here. 

\section{Exact Results}\label{sec:numerics}
\setcounter{equation}{0}

Thus far, we have established that in both the Fresnel and Fraunhofer regions no kinematic singularities occur in the localised matrix element $M$ given in~\refs{eq:matrix_element_def}. In addition, we have examined how the maximum of $|M|$ depends on the kinematic parameters, and also have shown that detection in the backwards direction is generally suppressed. In this section, we present typical numerical examples using the exact matrix element $M$ in~\refs{eq:matrix_element_def}, in order to analyse with greater accuracy its dependence on the momentum $\absp$, the distance $|\vecl|$, and the angle~$\theta$ (with $0\le \theta \le \pi$). The qualitative behaviour of the matrix element does not depend on the absolute scale of the parameters. Thus, we express all parameters in terms of $\dl$ as our basic unit of measurement, and fix the momentum of the exchanged $\Phi$ particle to have the value: $\qt=5/\dl$, chosen such that the effects we have been studying become readily visible in the figures we show in this section. The other parameters that appear in $M$ are varied independently, in order to showcase the various phenomena within the different near- and far-field regimes of interest, and at their interfacial regions. 

\subsection{Momentum Dependence}
\begin{figure}[t!]
		\centering \hspace*{-1.cm}
		\begin{subfigure}[b]{0.5\textwidth}
					\centering\includegraphics[width=1\textwidth ]{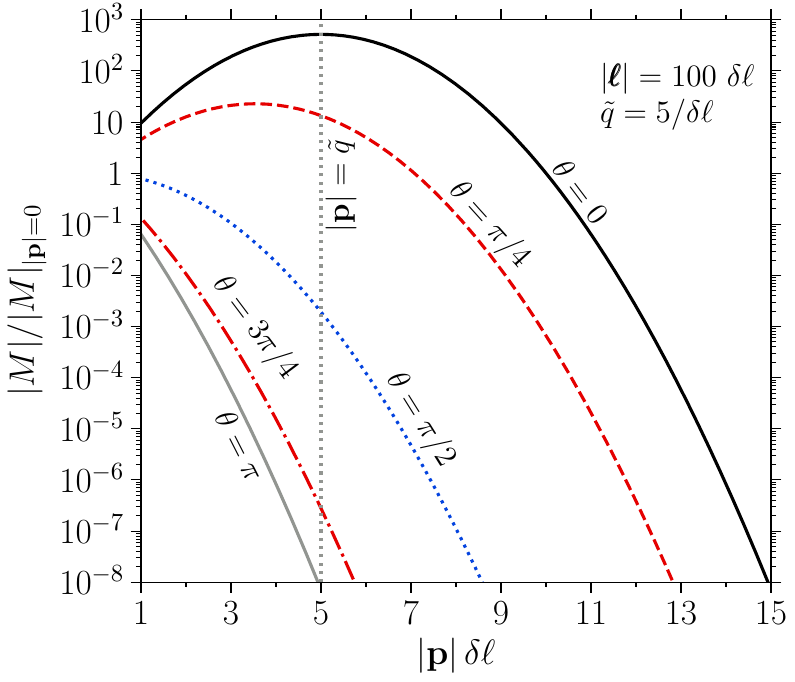}
					\caption{\label{fig:M_vs_p-far}}
		\end{subfigure}%
		\begin{subfigure}[b]{0.5\textwidth}
				\centering\includegraphics[width=1\textwidth ]{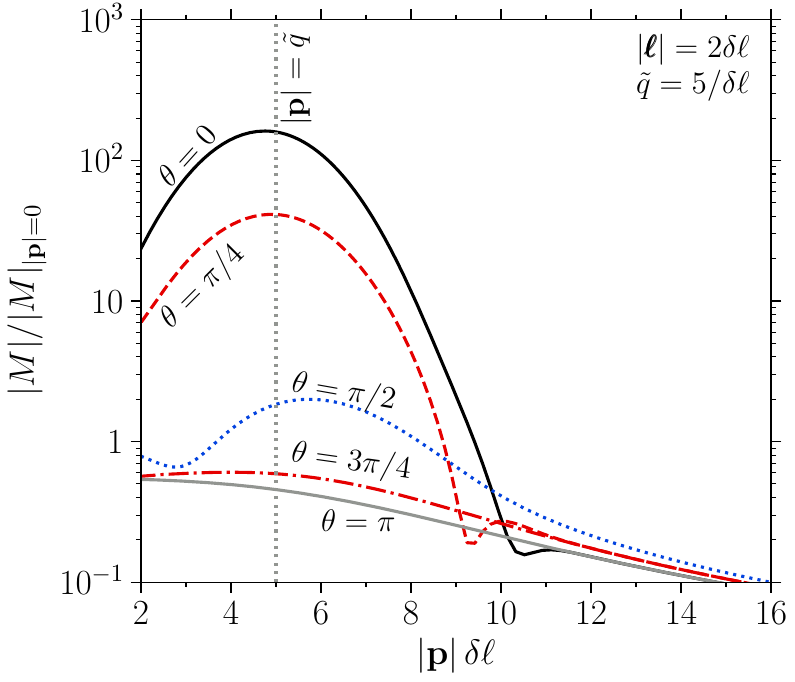}
				\caption{\label{fig:M_vs_p-near}}
    	\end{subfigure}%
		\caption{
    		{\bf (a)} The ratio $|M|/|M|_{\absp=0}$ as a function of $\absp$ for different azimuthal angles $\theta$ in the Fraunhofer zone, with $\absl=100\,\dl$, $\qt = 5/\dl$.
    		{\bf (b)} $|M|/|M|_{\absp=0}$ versus~$\absp$ for different angles $\theta$ in the Fresnel zone, with $\absl=2\,\dl$, $\qt = 5/\dl$. In both (a) and (b) panels, the gray vertical line corresponds to $\absp = \qt$.
		}
		\label{fig:M_vs_p}
\end{figure}

To start with, let us first consider the Fraunhofer region. In this region, the localised matrix element $M$ may be approximated as in~\refs{eq:M_far} and is exponentially dependent on the initial momentum~$\absp$. Like $|M|$ itself, its maximum also depends significantly on the angle $\theta$, defined by the vectors $\vecp$ and $\vecl$. In Figure~\ref{fig:M_vs_p-far}, we display a numerical example, which shows the value of the matrix element (over its value for $\absp=0$) in the far-field regime for a set of different angles. The distance between production and detection is set to $\absl=100 \, \dl$. As expected, the forward direction ($\theta=0$) corresponds to the maximum values of $|M|$, while larger observation angles $\theta$ give rise to lower values of $|M|$. The (global) maximum in the forward direction is obtained for $\pmax=\qt$,  while  $\pmax = \qt /\sqrt{2}$ when $\theta = \pi/4$. If observation occurs towards the backwards hemisphere ($\theta \geq \pi/2$), the  matrix element $M$ suffers a monotonous exponential suppression on $\absp$.

We now turn our attention to the Fresnel zone, in which $\absp \, \dl^2 > \absl$. To this end,  we show in \Figs{fig:M_vs_p-near} how $|M|/|M|_{\absp=0}$ changes with the momentum $\absp$ in this zone. In this near-field region, the matrix element exhibits a more complicated dependence on $\vecp$ and $\theta$. This is evident by the different forms we obtained in Section~3.2 for Subregions~\textbf{I},~\textbf{II}, and~\textbf{III} [cf.~\refs{eq:M_sub-I,eq:M_sub-II,eq:M_sub-III}].  Because of the specific choice of the spatial parameter, $\absl = 2\,\dl$, the three Fresnel subregions depend strongly on the observation angle~$\theta$.
For $\big|\absp - \qt\big|\dl \gg {\rm max}\big( 1,\;|\hatp\cdot\vecl|/\dl\big)$, all angles fall in Subregion~\textbf{I}. To be more precise, we observe that as $\absp$ surpasses $\qt$ all lines converge to the asymptotic curve of $|M| \propto 1/\absp^2$, as expected from~\refs{eq:M_sub-I}. 
If the mediator happens to be kinematically close to its mass shell, i.e.~when $\big|\absp - \qt\big| \, \dl \ll 1$, according to the approximate matrix elements~\refs{eq:M_sub-II,eq:M_sub-III}, we expect to find some maxima at initial momentum both above and below $\qt$. 
In the perpendicular direction $(\theta =\pi/2)$, if the resonant condition, $\absp \simeq \qt$, is satisfied, the observation vertex is always in Subregion~\textbf{II}. Hence, as $|\hatp \times \vecl|>\dl$, the maximum occurs when $\pmax \gtrsim \qt$ [cf.~\refs{eq:pmax_sub-II}].
Unlike $\theta =\pi/2$, the angles $\theta = 0$ and $\theta = \pi$ are entirely in Subregion~\textbf{III}, when $\absp \simeq \qt$. As $\absl = 2\,\dl$, the projection of $\vecl$ on $\hat\vecp$ is not well beyond the interaction radius, $\dl$.  Therefore, the estimate of the matrix element in  Subregion~\textbf{III} [cf.~\refs{eq:M_sub-III}] may not be accurate. Nevertheless, in the forward direction, we observe that $\pmax$ is close to the position of the maximum of the matrix element~$|M|$ as  approximated in~\refs{eq:M_sub-III}, \ie $\pmax \simeq \qt -2/(\qt\,\dl^2) \simeq 4.6/\dl$. On the other hand, for $\theta=\pi$, the maximum of $|M|$ in~\refs{eq:M_sub-III} occurs at $\pmax \simeq \qt - 2\absl^2/(\qt\dl^4) \simeq 3.4/\dl$. This results in $\big|\pmax -\qt\big|\,\dl^2 \simeq 1.6\,\dl \sim |\hat\vecp \cdot \vecl|_{\rm *}$, which indicates that this estimate may not be applicable in this case. Indeed, as shown in \Figs{fig:M_vs_p-near}, a numerical evaluation reveals a continuous decrease of the exact matrix element~\refs{eq:matrix_element_def} as $\absp$ increases.
Directions with $\theta = \pi/4$ and $\theta = 3\pi/4$ turn out to be close to the boundary between Subregions~\textbf{II} and~\textbf{III}, and have maxima at $\pmax \lesssim \qt$. In general, we can see that, apart from the successful regularization of the singularity at $\absp = \qt$, the matrix element exhibits distinguishable qualitative behaviour at different zones. This may be exploited by experiments, in searches for new particles as well as to study other potential implications of this formalism.

\begin{figure}[t!]
		\centering\includegraphics[width=0.5\textwidth ]{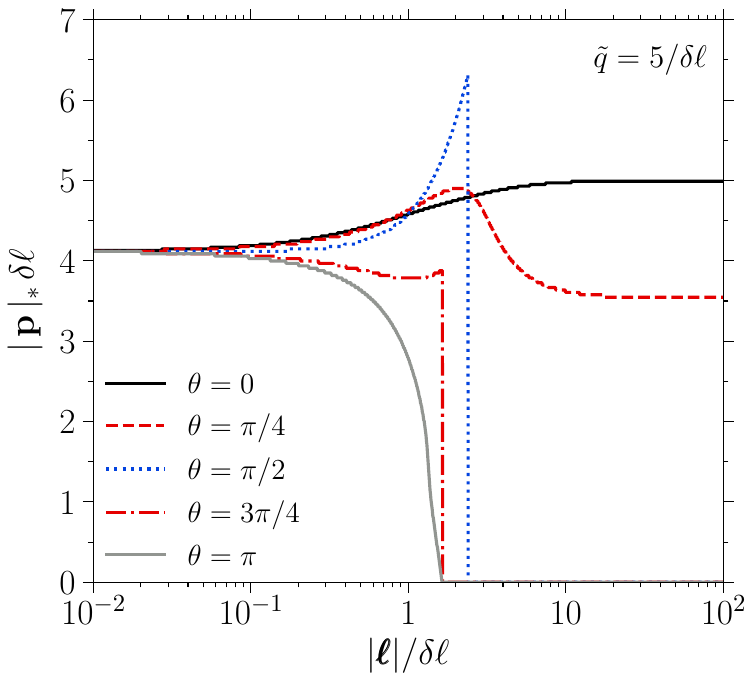}
		\caption{
		    The characteristic momentum $\pmax$, for which the maximum value of $|M|$ is attained, as a function of the mean distance $\absl$, for different angles~$\theta$, with $\qt = 5/\dl$. In the near-field region, in which $\absl \ll \absp \, \dl^2$, all detection angles $\theta$ converge to the same $\pmax$ value, which does not coincide with the usual OS momentum $\qt$. In the far-field region, where $\absl \gg \absp \, \dl^2$, $\pmax$ is highly $\theta$-dependent. In the forward direction, $\pmax = \qt$ as expected, while for $\theta>0$ $\pmax < \qt$. For angles $\theta > \pi/2$, $\pmax \to 0$, since the matrix element $|M|$ decreases monotonically with $\absp$. }
		\label{fig:p-res-shift}
\end{figure}
The aforementioned shift of $\pmax$ for various distances and angles is illustrated in~\Figs{fig:p-res-shift}. To be specific, \Figs{fig:p-res-shift} shows $\pmax$ as a function of $\absl$, for discrete choices of directions between $\theta = 0$ and  $\theta = \pi$. The values of $\pmax$ in the Fraunhofer zone agree with~\refs{eq:M_far}. That is, $\pmax = 0$ for $\theta \geq \pi/2$, $\pmax = \qt/\sqrt{2}$ for $\theta = \pi/4$, and $\pmax = \qt$ for $\theta = 0$.
Because of the assumed values of the input parameters, the resulting $\pmax$ cannot be estimated by~\refs{eq:M_sub-I,eq:M_sub-II,eq:M_sub-III}, for a wide range of~$\absl$ values in the Fresnel zone. However, \Figs{fig:p-res-shift} still reflects the behaviour expected from these estimates. In particular, as $\absl/\dl \to 0$, we expect that the maximum to occur at $\pmax \lesssim \qt$, as the matrix element is described by~\refs{eq:M_sub-II}, as long as $\big|\pmax -\qt\big|\dl \ll 1$. 
At greater distances, $\pmax$ can be both below and above $\qt$. Consider, for example, the numerical estimates of $\absp_*$, for $\theta=\pi/4$ in~\Figs{fig:p-res-shift}. As the distance between the production and detection increases, the corresponding Fresnel subregion changes from~\textbf{II} to~\textbf{III}. This causes $\pmax$ to increase between $\absl \ll \dl$ and  $\absl \simeq 2\,\dl$. As the distance $\absl$ is getting even larger, $\pmax$ moves towards its value found in the Fraunhofer region, \ie it falls to $\pmax \simeq \qt/\sqrt{2}$. This results in the maximum we observe in \Figs{fig:p-res-shift} around $\absl = 2\,\dl$.
In the perpendicular direction ($\theta=\pi/2$), if the resonant condition ($\big|\pmax - \qt\big|\dl \ll 1$) is satisfied, observation occurs in Subregion~\textbf{II} regardless of the distance~$\absl$, as long as $\absl \lesssim \pmax \, \dl^2$. According to~\refs{eq:pmax_sub-II}, this means that $\pmax$ starts lower than $\qt$, and increases as $|\vecp \times \vecl|$ surpasses $\dl$. Once $\absl$ approaches the boundary with the Fraunhofer zone, $\pmax$ drops to $\pmax=0$, around $\absl = 2.5 \, \dl$.

\subsection{Spatial Dependence}

One important aspect of LSMT under consideration is its introduction of an explicit dependence on the average distance vector $\vecl$ between the production and detection vertices. This explicit radial dependence has been used to model neutrino oscillations~\cite{Ioannisian:1998ch}, but it can also be used for other studies, such as displaced vertex searches~\cite{LHCb:2014osd,Bondarenko:2019tss} for long-lived particles. Therefore, it is worth examining the predictions derived from the exact matrix element $M$ in~\refs{eq:matrix_element_def} in the Fraunhofer zone, the three Fresnel subregions, and all their interfaces. 

\begin{figure}[t!]
	\centering \hspace*{-1.cm}
	\begin{subfigure}[b]{0.5\textwidth}
		\centering\includegraphics[width=1\textwidth ]{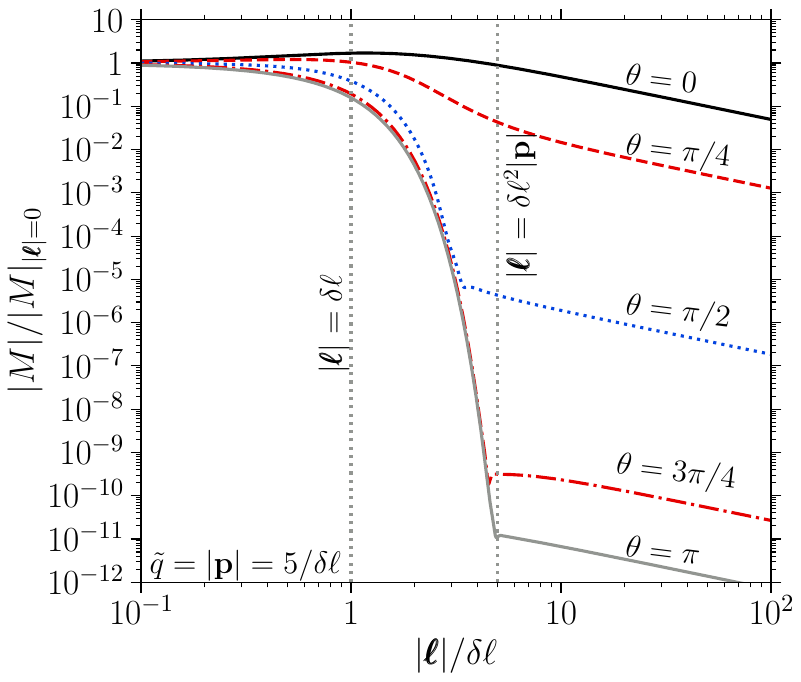}
		\caption{\label{fig:M_vs_l_on-shell}}
	\end{subfigure}%
	\begin{subfigure}[b]{0.5\textwidth}
		\centering\includegraphics[width=1\textwidth ]{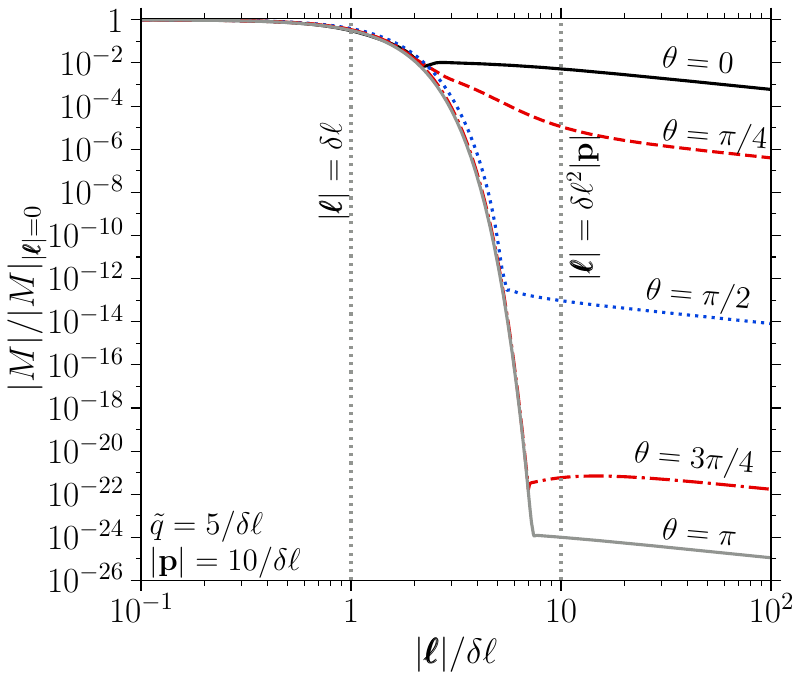}
		\caption{\label{fig:M_vs_l_off-shell}}
	\end{subfigure}%
	\caption{
    	{\bf (a)} The ratio $|M|/|M|_{\absl=0}$ versus distance $\absl$, for $\absp=\qt=5/\dl$ and discrete choices of the angle $\theta$ between $0$ and $\pi$. 
    	{\bf (b)} The same as in the left frame~(a), but with  $\absp=10/\dl$. 	
	}
	\label{fig:M_vs_l}
\end{figure}

To illustrate the spatial dependence of the exact matrix element $M$~in~\refs{eq:matrix_element_def}, we show in \Figs{fig:M_vs_l_on-shell} the ratio $|M|/|M|_{\absl=0}$ as a function of~$\absl$ in the OS region where~${\absp =\qt}$, for selected values of $\theta$ between $0$ and $\pi$. 
As can be seen from this figure for $\theta = 0$ (in black), there is a maximum at a location away from the source. This implies a greater flux of outgoing particles in the forward direction. This is a distinct prediction that originates from our localised S-matrix amplitude and might well be tested in dedicated experiments. 
There is a significant dependence on the direction of $\vecl$, and as $\theta$ gets larger, the matrix element $M$ displays no local maximum.  In particular, for $\theta \geq \pi /2$, the exact $M$ decreases exponentially between $\absl \simeq \dl$ and $\absl \simeq \absp \, \dl^2$ (the vertical gray lines). This means that propagation towards the backwards hemisphere~({$\theta \ge \pi/2$}) is suppressed, which is consistent with our findings  for the near- and far-field approximations in~\refs{eq:M_far,eq:M_sub-II,eq:M_sub-III}. Finally, it is interesting to notice that in the Fraunhofer region ($\absl \gg \absp\,\dl^2$), the matrix element evaluated at a point in the forward direction ($\theta = 0$) is more than ten orders of magnitude larger than its value at an equidistant point, lying in the backwards direction~($\theta =\pi$).  

In \Figs{fig:M_vs_l_off-shell} we now show the spatial dependence of the exact matrix element in~\refs{eq:matrix_element_def} for an  off-shell 
kinematic configuration, with $\absp=10/\dl$, while the rest of the parameters are as in~\Figs{fig:M_vs_l_on-shell}. We observe that the absence of a maximum away from the origin in any direction. Also, for $\absl \lesssim2\,\dl$, all directions result in similar values of $|M|$, as expected from~\refs{eq:M_sub-I}. This means that off-shell propagation close to the interaction area can occur in all directions with equal probability. However, at greater distances, the forward direction is preferred, as $|M|$ falls off exponentially for larger angles. Like in the on-shell case, at $\absl \gg \absp \, \dl^2$, all $\theta$ angles predict a matrix element $|M| \propto 1/\absl$.     

\begin{figure}[t!]
		\centering \hspace*{-0.35cm}
		\begin{subfigure}[b]{0.5\textwidth}
					\centering\includegraphics[width=1\textwidth ]{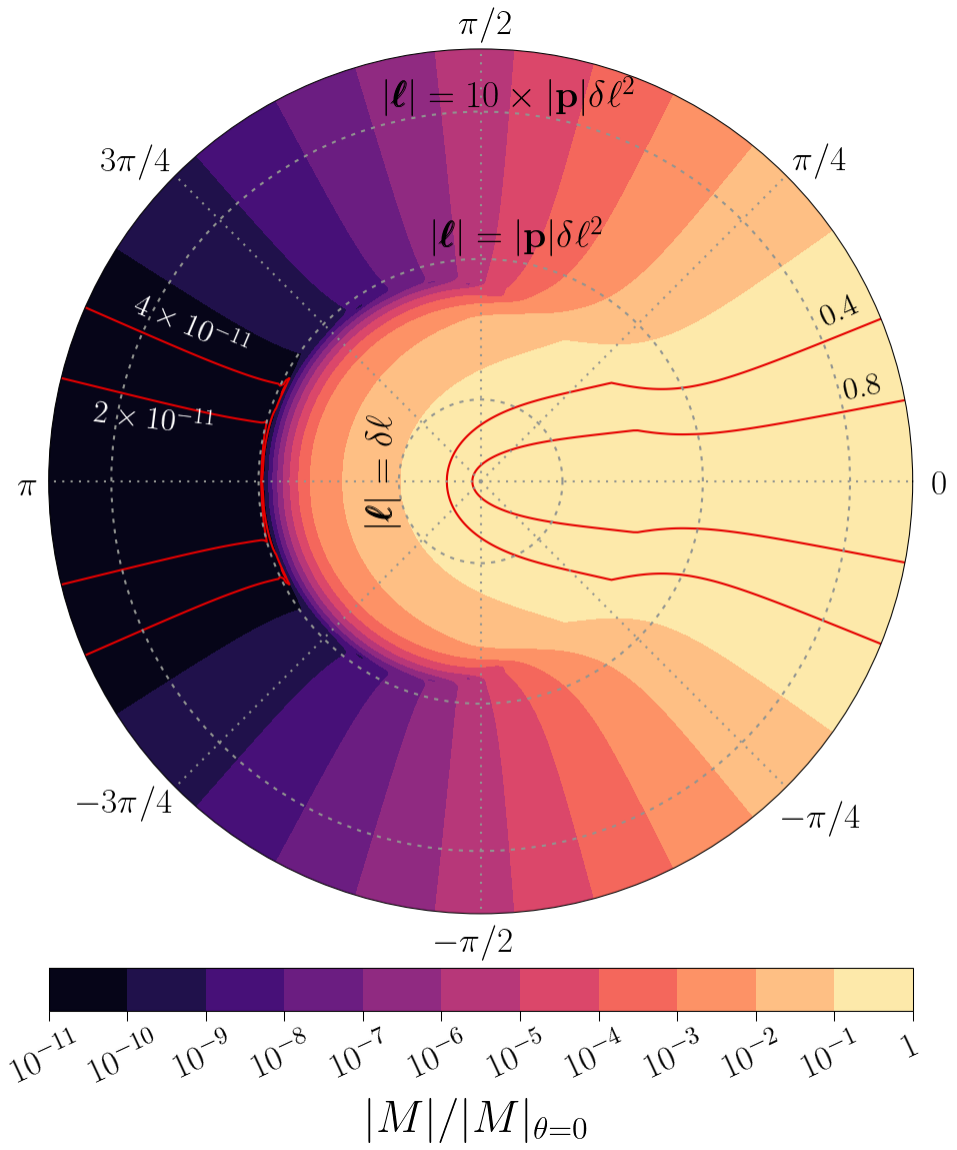}
					\caption{\label{fig:contour_norm_on-shell}}
		\end{subfigure}%
		\hspace*{0.2cm}
		\begin{subfigure}[b]{0.5\textwidth}
				\centering\includegraphics[width=1\textwidth ]{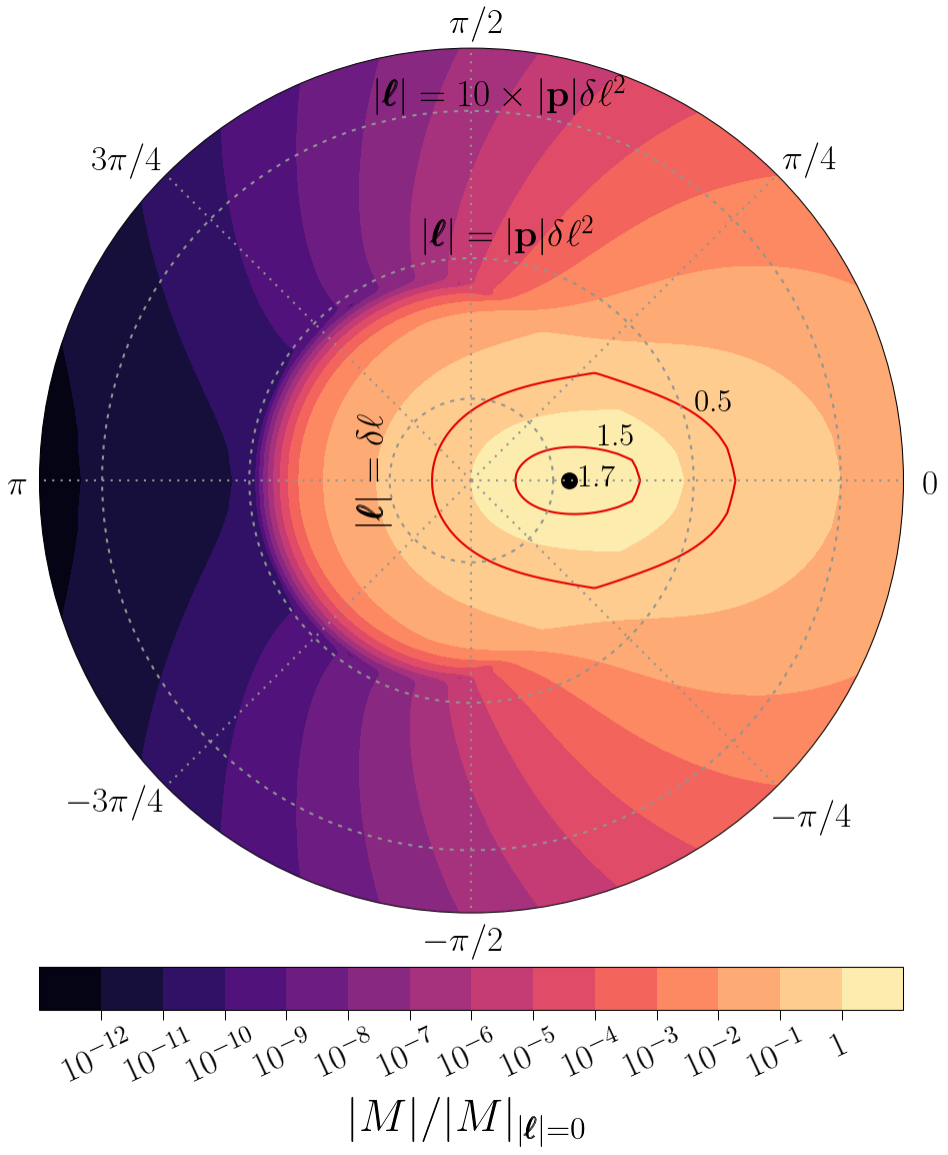}
				\caption{\label{fig:contour_on-shell}}
    	\end{subfigure}%
		\caption{
		    {\bf (a)} Numerical estimates of the exact matrix element $|M|$ in~(\ref{eq:matrix_element_def}) normalised by its value in the forward direction, $|M|_{\theta=0}$. The red contours show specific values for $|M|/|M|_{\theta=0} =0.8,\,0.4,\,4 \times 10^{-11},\text{ and }2 \times 10^{-11}$.
    		{\bf (b)} The same as in (a), but for $|M|$ normalised by its value $|M|_{\absl=0}$ at the origin $\absl=0$. The red contours delineate the curves on which the logarithm of the aforementioned ratio is $1.5$ and $0.5$. The black point  outside $\absl=\dl$ indicates the maximum of $|M|/|M|_{\absl=0}\simeq 1.7$.
    		In both panels (a) and (b), the values of the parameters are taken as in Figure~\ref{fig:M_vs_l_on-shell}. The gray circles show $\absl = \dl$,  $\absl = \absp \, \dl^2 $ (the boundary between the Fraunhofer and Fresnel regions), and $\absl=10 \, \absp \, \dl^2$. The various colours show the order of magnitude of the ratios: $|M|/|M|_{\theta=0}$ in (a) and $|M|/|M|_{\absl=0}$ in (b), for a given distance~$\absl$ and angle~$\theta$.
	    }
	\label{fig:contours_on-shell}
\end{figure}

\begin{figure}[t!]
		\centering \hspace*{-0.35cm}
		\begin{subfigure}[b]{0.5\textwidth}
					\centering\includegraphics[width=1\textwidth ]{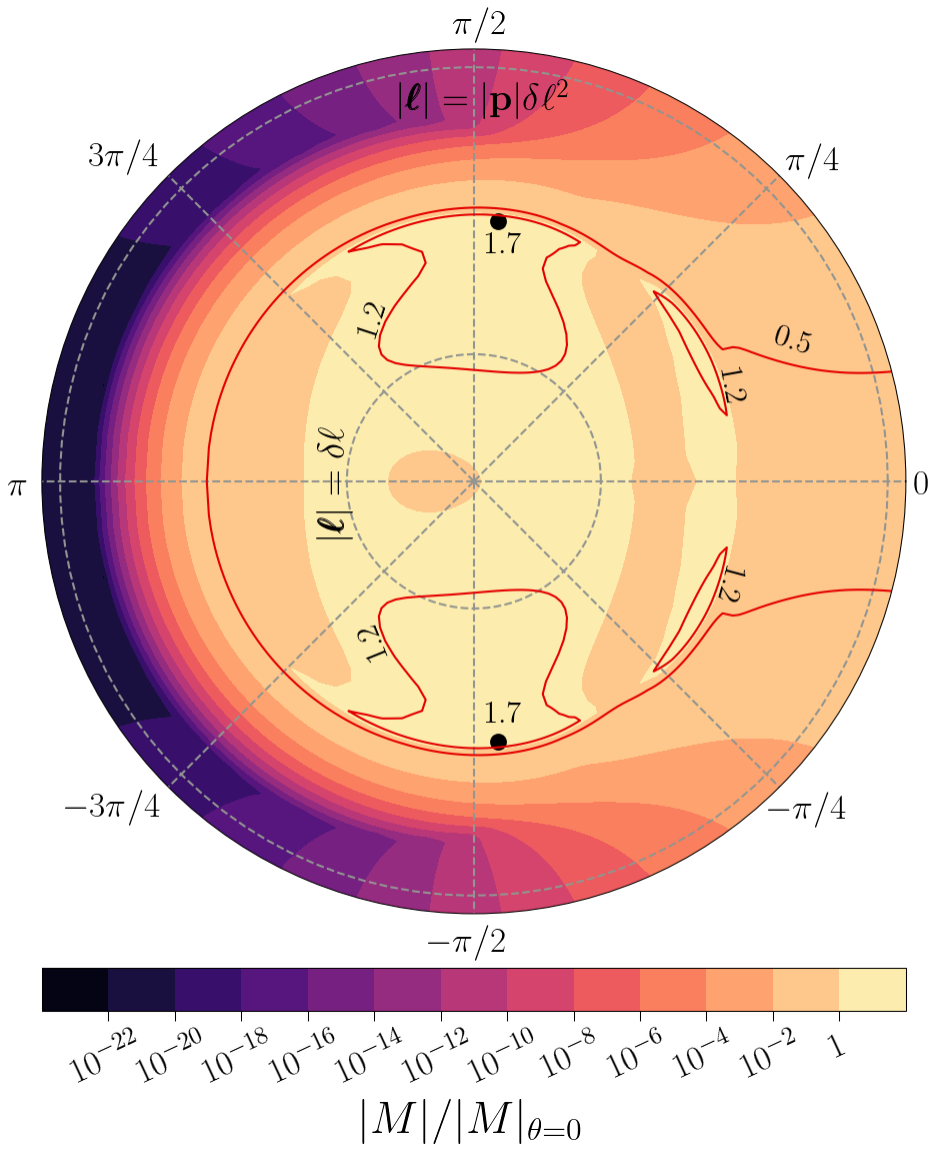}
					\caption{\label{fig:contour_norm_off-shell}}
		\end{subfigure}%
		\hspace*{0.2cm}
		\begin{subfigure}[b]{0.5\textwidth}
				\centering\includegraphics[width=1\textwidth ]{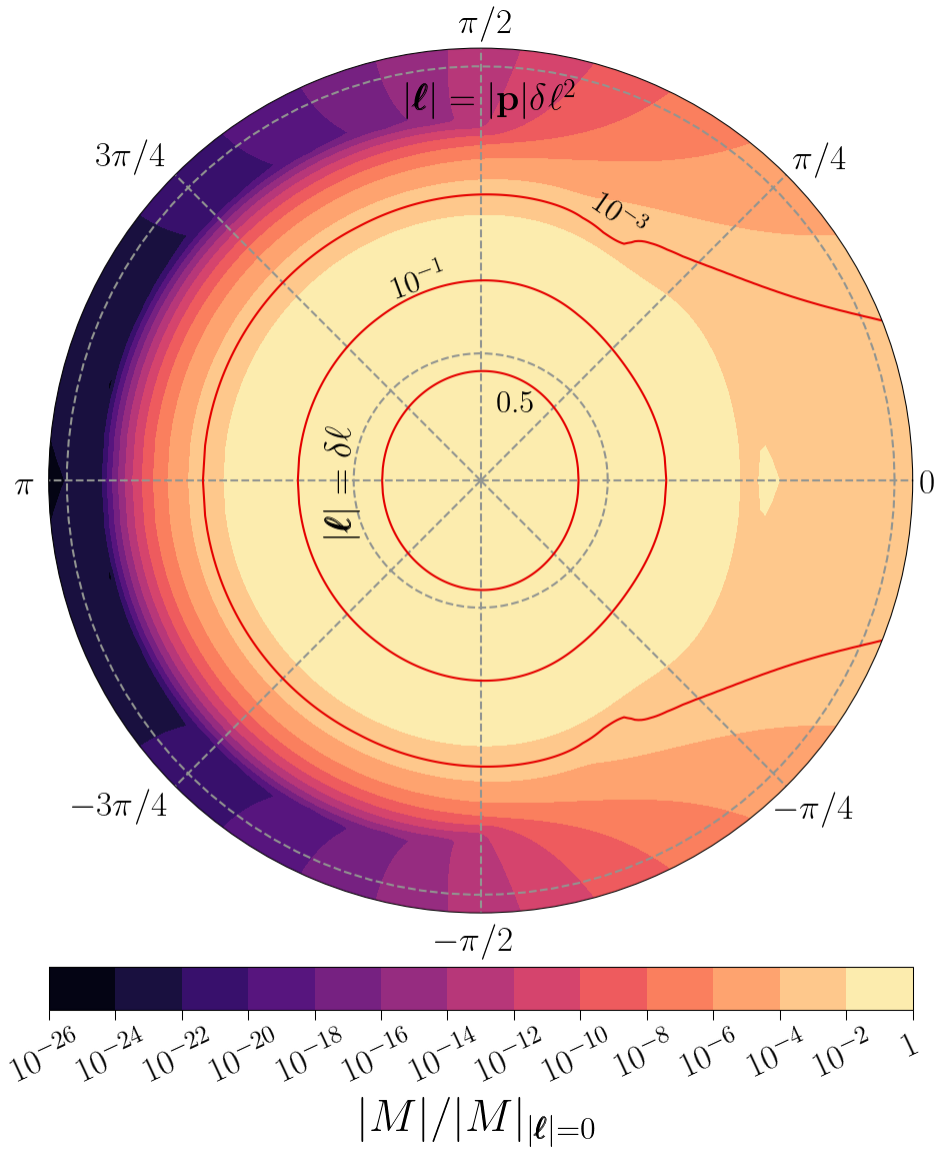}
				\caption{\label{fig:contour_off-shell}}
    	\end{subfigure}%
		\caption{ 
            {\bf (a)} The ratio $|M|/|M|_{\theta=0}$ in polar coordinates $(\absl,\, \theta)$ in the Fresnel zone, for the same input parameters as in Figure~\ref{fig:M_vs_l_off-shell}. The red curves correspond to $|M|/|M|_{\theta=0} = 0.5$ and $1.1$. The maximum value of this ratio is $|M|/|M|_{\theta=0} = 1.7$ (black dots), obtained approximately along the perpendicular direction $\theta \simeq \pi/2$.
            {\bf (b)} The ratio $|M|/|M|_{\theta=0}$ in the Fresnel region, for the same input parameters as in~(a). 
            The red curves correspond to $|M|/|M|_{\theta=0} = 0.5,\,0.1$, and $10^{-3}$. In contrast to the on-shell case, the maximum value of this ratio corresponds to $\absl \simeq 0$ in the forward direction.
        }
	\label{fig:contours_off-shell}
\end{figure}

In \Figs{fig:contours_on-shell,fig:contours_off-shell}, we present, as polar density plots, the radial and angular dependence of the exact matrix element $M$ in~\refs{eq:matrix_element_def} for on-shell and off-shell kinematic configurations of the mediator propagator.
More explicitly, in \Figs{fig:contour_norm_on-shell}, we show $|M|$ normalized with respect to its value at $\theta =0$ for $\absp = \qt = 5/\dl$. The radial parameter is $\absl$ with the three grey concentric circles indicating $\absl = \dl$, $\absp \, \dl^2$, and $10 \times \absp \, \dl^2$. The various colours represent the order of magnitude of $|M|/|M|_{\theta=0}$, and we explicitly show four curves (in red) with the values $|M|/|M|_{\theta=0} = 0.8,\,0.4,\,4\times10^{-11}$, and $2 \times 10^{-11}$. As expected, in the far-field region the radial dependence~$\absl$ cancels out as $|M|/|M|_{\theta=0} \simeq e^{\absp \qt \cos\theta /2}$. Furthermore, in the near-field region, the radial dependence cannot be factored out, as implied by~\refs{eq:M_sub-II} and~\eqref{eq:M_sub-III}. As a result, the spatial pattern in the Fresnel zone displays a strong angular dependence. 
Focusing on angles $\theta \lesssim \pi/4$, \eg~looking at the curve for $|M|/|M|_{\theta = 0} =0.8$,  both $\absl$ and~$\theta$ increase for $\absl > \dl$, in order to keep the ratio $|M|/|M|_{\theta = 0}$ constant. However, as $\absl$ approaches $\absp \, \dl^2$, $\theta$~starts decreasing until $\absl \simeq \absp \, \dl^2$. Thus, $|M|/|M|_{\theta = 0} =0.8$ has a non-trivial behaviour as observation moves from the near-field to the far-field zone. 

In \Figs{fig:contour_on-shell}, we display the norm of the matrix element over its value at the origin, $|M|/|M|_{\absl=0}$, for the same parameters as in~\Figs{fig:M_vs_l_on-shell}. The various colours represent the order of magnitude of $|M|/|M|_{\absl=0}$, along with the two curves (in red) for $|{M|/|M|_{\absl=0} = 1.5}$~and~$0.5$. We also indicate with a black dot the point where the global maximum occurs. This figure shows the overall $\absl$ and $\theta$ dependence of $|M|$. We observe that in the far-field regime, the behaviour of $|M|/|M|_{\absl=0}$ matches perfectly well with that predicted by~\refs{eq:M_far}. Like in~\Figs{fig:contour_norm_on-shell}, there is an effective boundary at $\absl \simeq \absp \, \dl^2$ that severely restricts propagation in the backwards hemisphere ($\theta \ge \pi/2$). \Figs{fig:contour_on-shell} also shows how the maximum observed in \Figs{fig:M_vs_l_on-shell} changes for different angles and distances. We notice that $|M|$ only increases for $\theta<\pi/2$ and reaches a maximum indicated by a black dot at which $|M|/|M|_{\absl = 0} \simeq 1.7$. This maximum occurs at a distance marginally larger than $\dl$ at $\theta=0$.

%
Although an off-shell mediator will generate a  similar far-field pattern (for $\qt^2>0$), the Fresnel regime is quantitatively different. This is shown in \Figs{fig:contour_norm_off-shell}, where we compute $|M|/|M|_{\theta=0}$ for the same parameters as in~\Figs{fig:contour_norm_on-shell}, but for the off-shell point: ${\absp=2\qt}$. In~contrast to the OS case ($\absp = \qt$), this ratio exhibits maxima away from the origin, in the perpendicular direction~$(\theta = \pi/2)$, which are indicated symmetrically with two black dots. Interestingly, the ratio is larger than~$1$, even towards the backwards direction well within the Fresnel zone. However, close to the interface between the near- and far-field zone, there is a drastic exponential suppression when $\theta > \pi/2$, similar to the one we saw in~\Figs{fig:contour_norm_on-shell}. At larger distances, the matrix element falls off as $1/\absl$, in agreement with our expectations in the Fraunhofer regime [cf.~\eqref{eq:M_far}].    

\Figs{fig:contour_off-shell} shows the spatial profile of $|M|/|M|_{\absl=0}$ for an off-shell mediator with $\absp=2\qt$. We observe that there are significant differences with its on-shell counterpart in~\Figs{fig:contour_on-shell}. Specifically, the curves, for which the ratios $|M|/|M|_{\absl=0}$ are being kept constant, are almost independent of the angle well within the Fresnel zone. This property is also reflected in the approximation~\refs{eq:M_sub-I}, as well as in \Figs{fig:M_vs_l_off-shell}. Thus, close to the source, no preferred direction exists and propagation happens at all angles with almost equal probabilities. As $\absl$ approaches $\absp \, \dl^2$, the forward direction becomes more favourable, whereas the matrix element for angles $\theta \gtrsim \pi/2$ falls off exponentially, in line with our numerical estimates in~\Figs{fig:M_vs_l_off-shell}. 

In summary, we find that the numerical estimates presented here
by utilising the exact localised matrix element~$M$ stated in~\eqref{eq:matrix_element_def} give firm support to the validity of the more intuitive Fresnel and Fraunhofer approximations discussed in Section~\ref{sec:approx}.

\subsection{Possible Experimental Probes}
The approach we have been studying in this article can be applied to a number of experiments. For instance, in $\mu^+ \, \mu^-$ colliders the LSMT will provide a regularization of $t$-channel singularities~\cite{Nowakowski:1993iu,Ginzburg:1995bc,Kotkin:1992bj,Melnikov:1996iu,Melnikov:1996na,Grzadkowski:2021kgi} by means of QM uncertainties. Notably, the LSMT can be used to explain the spatial pattern of diffraction well beyond the realm of classical electrodynamics. As an example,  we discuss in this subsection neutrino experiments, like the currently projected long-baseline neutrino experiment DUNE~\cite{DUNE:2020lwj}, which can be more challenging. 

In such neutrino experiments like DUNE, a proton beam collides with a target to produce charged pions, $\pi^\pm$. These pions pass through a series of magnets that cause them to follow more converging trajectories. Then, the so-collimated beam of pions decay into neutrinos (and leptons), thus creating a neutrino beam. Several detectors along the path of the neutrino beam measure the ``neutrino flux''. 
Strictly speaking, in the LSMT neutrinos are considered to be mediators of the interactions between the pion beam and the detector.\footnote{These interactions proceed through the $t$-channel which can in principle exhibit singularities along with the neutrino oscillations. The LSMT can incorporate consistently both the regularisation of $t$-channel singularities and neutrino oscillations in space.} This in turn means that the neutrinos are virtual particles and so they are not directly observed. 
For instance, in the far-field region described by the matrix element~\refs{eq:M_far}, neutrinos could be interpreted as spherical waves propagating with momentum of magnitude $\qt$, which means that the expected scaling $M\sim 1/ \absl$ (i.e.~inverse squared law) is justified. This semi-classical interpretation breaks down for distances close to or inside the Fresnel zone, as the matrix element in the several subregions takes on different forms.
A typical event is usually registered as energy deposition to the detector due to its interaction with the pion beam. According to the LSMT, the characteristics of this interaction, such as its angular distribution, is a consequence of a flux of virtual neutrinos, with event rates as predicted by the amplitude~\refs{eq:amplitude_k=p_final}. 

It is worth mentioning that the angular distribution of number of events for different pion energies can be measured by utilising the technology of Liquid Argon Time Projection Chambers (LArTPCs). Particle detectors based on LArTPCs have excellent scintillation properties~\cite{Majumdar:2021llu}, as well as unique capabilities of measuring within the flux both the position and the direction of the propagating neutrinos at different distances from the source. An experimental project being in progress is the short-baseline neutrino~(SBN) programme at Fermilab that consists of ICARUS~\cite{ICARUS:2004wqc}, MicroBooNE~\cite{MicroBooNE:2016pwy}, and the short-baseline near detector (SBND)~\cite{Tufanli:2016hyo}. 

For the aforementioned experiments, the angular distribution of the number of events may be evaluated as    
\begin{equation}
    \dfrac{d N}{ d\Omega} = \int dE_{\pi^\pm} \, \dfrac{d\sigma_{\pi^\pm}}{d\Omega} \ \dfrac{d\mathcal{L}_{\pi^\pm}}{dE_{\pi^\pm}} \;.
    \label{eq:N_dist}
\end{equation}
Here, $\mathcal{L}_{\pi^\pm}$ is the luminosity of the pion beam, $E_{\pi^\pm}$ is the energy of the charged pions, and $\sigma_{\pi^\pm}$ is the total cross section of all interactions between the pion beam and the detector. The luminosity at different pion energies can be measured in similar experiments or extracted from simulations, e.g.~by making use of the {\tt Geant4} toolkit~\cite{Ivanchenko:2003xp}. In such a setup, the cross section can be calculated using the matrix element~\refs{eq:matrix_element_def}. The momentum $\vecp$ that determines the value of $\sigma_{\pi^{\pm}}$ can be found by precisely measuring $\veck$, which is the momentum difference between the initial and final momenta of the particles in the detector. For instance, if the pion beam-detector interaction results in a recoiled nucleon inside the detector, then $\veck$ becomes the final-state momentum of the nucleon in the laboratory frame.

In an idealised experiment, measuring and comparing against LSMT predictions should in principle be straightforward. However, there can be several experimental and theoretical challenges under more realistic conditions. For instance, in LArTPCs\- there are other processes that can take place, beyond $2 \to 2$ collisions, which need to be taken into account consistently within the LSMT framework. Since both statistical and quantum uncertainties affect measurements,  analyses analogous to the studies in~\cite{Campagne:1997fu,Kiers:1997pe,Ioannisian:1998ch,Beuthe:2001rc} will provide more accurate results. 
Another possible complication may arise from considering a non-spherically symmetric spatial smearing, which is expected to make the computation of the corresponding amplitudes more difficult. In the absence of an analytic form, numerical methods can be used to compute these amplitudes~\cite{Pittau:2021jbs}. 
Such analyses are beyond the scope of our paper, since our main concern here is to lay the foundations for a consistent formulation of LSMT. 

We note that the exact value of $\dl$ cannot be obtained from first principles and should be inferred from the experiment. For instance, performing experiments on similarly prepared pion beams can help us estimate or impose bounds on $\dl$. Such bounds will enable us to place detectors in the Fraunhofer region, which can be used to determine an accurate value of $\dl$ that corresponds to each single-energy band of the initial-state pion beam.

Estimating $\dl$ from the morphology of the pion beam will have significant implications for the predictions obtained from the LSMT. Let us, for example, consider a pion beam for which the individual pions exhibit a QM uncertainty $\dl \sim 10^{-5}~{\rm cm}$. In addition, let us also assume that the magnitude of the average momentum of the pion beam is $\absp \sim 10~\GeV$. In such an experimental setting, the interface between the Fresnel and Fraunhofer zones, $\absl = \dl^2 \, \absp$, extends up to $\absl  \sim 1~{\rm km}$. This will enable us to probe all subregions of the near-field regime by placing conveniently the detectors at distances $\absl \lesssim 1~{\rm km}$. Such arrangements will provide another possible experimental probe for testing the validity of our LSMT.

\section{Summary and Future Directions}\label{sec:summary}
\setcounter{equation}{0}

Non-locality, as expected to originate from the Feynman propagator, is an inherent property of~QM and plays an instrumental role in understanding several non-local phenomena in many applications of modern quantum theory, ranging from simple two-particle quantum-entangled systems, like those that occur in an EPR experiment~\cite{Einstein:1935rr}, to more complex situations in quantum information and quantum techno\-logy~\cite{Alonso:2022oot}. Here, our aim was to extend this notion of non-locality to the standard S-matrix of QFT. In particular, we put forward an S-matrix theory in which each particle interaction in a scattering process is taken to be localised in a volume of finite size. For brevity, we called such a theory the Localised S-Matrix Theory (LSMT). Evidently, such an LSMT assumes its standard S-matrix form, when the infinite spread limit in the localisation of all interactions is considered.

To gain insight into the formalism of the LSMT, we have considered a simple $2\to 2$ scattering process within an analytically solvable QFT model that was previously discussed in~\cite{Ioannisian:1998ch}. This solvable QFT model is based on two working hypotheses. First, we have taken the QM uncertainty in time, $\delta t$, to be much bigger than the combined QM
uncertainty $\dl$ of the detector and the source. In fact, we have worked in the limit of~${\delta t \to \infty}$, which in turn implies that the (mean) energy is conserved at each interaction vertex of the scattering process. Second, we have assumed that both the production and detection points of interaction have spatial spreads with spherical Gaussian form. The latter hypothesis enables us to carry out most of the complex integrations that we encounter, and so arrive at an analytic result that only depends on well-tabulated complementary error functions with complex arguments. In spite of the above assumptions, we should expect that the results presented here for the different near- and far-field zones will still be generically valid, up to obvious amendments, for other scenarios with QM localisations that go beyond the spherical approximation considered here. 

In the context of a solvable QFT model discussed earlier in~\cite{Ioannisian:1998ch}, we have derived several analytic approximations of the localised S-matrix amplitude for detection regions that are either quite close to the source or very far from it. Adopting a termino\-logy known from light diffraction in classical optics, we called these two regions interchangeably the near-field and far-field zones, or the Fresnel and Fraunhofer regions. The Fresnel (near-field) zone is confined to distances $\absl$ from the source that lie in the interval, $0 \le \absl \lesssim \absp\,\dl^2$, where $\vecp$ is the net three-momentum of all particles in the initial or final state of the process. Instead, the Fraunhofer (far-field) region characterises the region far from the source, for which $\absl \gg \absp\,\dl^2$.  

We have found that the Fresnel zone may be subdivided into three subregions according to the values of the two dimensionless quantities, $|\hatp\cdot \vecl |/\dl$ and $\big|\absp - \qt \big|\,\dl$. A more detailed description of Subregions {\bf I}, {\bf II} and {\bf III} is given in~Table~\ref{tab:fresnel_Subregion}. For all these three Fresnel subregions, we observed that the on-shell transition amplitude $M$ does \emph{not} fall off as $1/\absl$ as a function of the distance~$\absl$ between the source and the detector, thereby confirming the non-dispersive, plane-wave behaviour of $M$ in the forward direction, in agreement with earlier observations made first in~\cite{Ioannisian:1998ch}, and subsequently in~\cite{Beuthe:2001rc,Naumov:2013bea,Naumov:2022kwz} in different settings. Remarkably enough, in the same forward direction of propagation, we have observed a novel focusing phenomenon manifesting itself with the appearance of a small area where the magnitude $|M|$ of the transition amplitude can be higher than its value at the origin, where~${\absl=0}$. As expected, in the Fraunhofer region, we recover the usual $1/\absl$ reduction of $|M|$. In both the near- and far-field regions, we have confirmed the phenomenon of oscillations if the mediators form a mixed system of particles, as is the case, for example, for neutrino oscillations. 

Another novelty of the present study is the analysis of the transition amplitude~$M$ beyond the forward direction of propagation, as a function of the angle $\theta$ defined by the average distance vector~$\vecl$ and the total three-momentum vector ${\bf p}$ of the particles in the initial state. An~important finding of such an analysis was the observation that in the backwards direction ($\theta = \pi$), the amplitude $M$ is extremely suppressed. One may therefore conclude that the Feynman propagator provides the necessary ``quantum obliquity factor'' to suppress the propagation of on-shell particles in the backwards direction. We must emphasize here that this desirable property of~$M$ is achieved without the need to impose certain boundary conditions on the system. In this way, the LSMT can provide a quantum field-theoretic explanation for the origin of the obliquity factor in diffractive optics. In the same vein, it is appealing to suggest that the analytic result for $M$ (which depends on complexified error functions) represents an analytic QFT extension of the famous Euler--Cornu spiral~\cite{Optics_1999} in classical optics to the complete off-shell region of particle propagation.  

In realistic situations, we expect that the temporal and spatial QM uncertainties due to finite space-time volume
effects on a localised S-matrix amplitude,~$M$, will depend on the experimental setup, including the preparation and detection of the initial and final states. In addition to the coherent QM uncertainties, one must therefore include \emph{incoherent} statistical uncertainties to be added at the squared amplitude level, $|M|^2$, along with phase-space and other classical resolution effects~\cite{Giunti:1993se,Beuthe:2001rc,Akhmedov:2009rb,Naumov:2020yyv,Cheng:2022lys,Naumov:2022kwz}. In this context, 
the LSMT offers an important element in a holistic construction of a more elaborate multi-local Wigner function~\cite{Wigner:1932eb,Hillery:1983ms} which may include all possible uncertainties for all realistic experimental settings. Hence, as well as both short and long baseline neutrino experiments, future high-energy colliders have the potential to probe many of the predictions resulting from such an LSMT. For instance, one may exploit the crossing symmetry of the localised S-matrix amplitude to regulate the notorious $t$-channel singularities at $\mu^+\mu^-$ colliders~\cite{Nowakowski:1993iu,Ginzburg:1995bc,Kotkin:1992bj,Melnikov:1996iu,Melnikov:1996na,Grzadkowski:2021kgi}. Other applications of the LSMT may include spatial analyses of parton showering and displaced vertices during the hadronization process at high-energy colliders like the LHC~\cite{LHCb:2014osd,Bondarenko:2019tss}. We envisage that such analyses\- might also lead to improved interpretation of data from $B$-meson observables at~the~LHCb.


In this paper we only laid out the foundations for an analytic LSMT. However, further work must be done if we wish to go beyond the Born approximation. For example, for the $2\to 2$ process under study, we expect that box contributions to the localised transition amplitude $M$ will decay exponentially faster with increasing distance $\absl$ from the source than  the one-particle-reducible propagator effects. In this way, a physical separation between the irreducible (box) and reducible (self-energy) loop diagrams may be possible, thus enabling a better understanding of S-matrix diagrammatic approaches like those based on the pinch technique~\cite{Binosi:2009qm}. On the other hand, apart from scalar mediators that we have analysed here in a solvable QFT model, it should be straightforward to generalise LSMT and include localised exchange graphs with fermions and gauge bosons. We~shall return to address some of the issues mentioned above in a future study. 

\bigskip
\subsection*{Acknowledgements}

We wish to thank Cumrun Vafa and Vassilis Spanos for remarks and useful comments at the \emph{XXIX International Conference on Supersymmetry and Unification of Fundamental Inter actions} 
(SUSY~2022, 27~June~--~2~July~2022, University of Ioannina, Greece), where part of this work was first presented.
We also thank Bohdan Grzadkowski and Michal Iglicki for clarifying discussions with regards to~\cite{Grzadkowski:2021kgi}, as well as Stefan Söldner-Rembold and Anyssa Navrer-Agasson for discussions concerning the SBN programme at Fermilab.
The work of AP and DK is supported in part by the Lancaster-Manchester-Sheffield Consortium for Fundamental Physics, under STFC Research Grant ST/T001038/1.

\newpage
\setcounter{section}{0}
\section*{Appendix}
\appendix

\renewcommand{\theequation}{\Alph{section}.\arabic{equation}}
\setcounter{equation}{0}  

\section{Calculation of the Localised S-Matrix Amplitude}\label{app:amplitude}
\setcounter{equation}{0}

Here we will present the main steps that we followed to derive the analytic expression~\eqref{eq:amplitude_final} for the amplitude $T_{\rm L}$ pertaining to the process ${S_1\chi_1\to \Phi^*(q)\to S_2\chi_2}$.

To start with, we first note that integration over the time coordinates $x^0,\, y^0$ and $q^0$ in~\refs{eq:amplitude_init} can be carried out using the usual definition of $\delta$-functions. More explicitly, we have
\begin{equation}
	\dint dx^0 \, dy^0 \dfrac{dq^0}{2\pi} e^{-i p^0 \, x^0} \, e^{i k^0 \, y^0} \, e^{i q^0 \, (x^0-y^0)} G(q^0,\bvec q)= 2\pi \  \delta(p^0 - k^0) \ G(p^0,\bvec q)\;,
	\label{eq:time-int}
\end{equation}
where $G(p^0,\bvec q)$ is some analytic function with respect to $p^0$. Note that the appearance of $\delta(p^0 - k^0)$ is a consequence of energy conservation in the infinite limit of 
time uncertainties, that is for $\delta x^0,\, \delta y^0 \to \infty$. Making use of~\eqref{eq:time-int}, the localised amplitude reads
\begin{align}
	T_{\rm L}(p,k; \xb,\yb,\dx,\dy) = & -2\pi \, \delta(p^0 - k^0) \ \lambda \, g \dint d^3 \bvec x \, d^3 \bvec y \ e^{-(\bvec{x}-\xb)^2/\dx^2 } \,  e^{-(\bvec{y}-\yb)^2/\dy^2 }  \nonumber \\
	&\times\, e^{i \vecp \cdot \bvec x - i \veck \cdot \bvec y} \dint \dfrac{d^3 \bvec q}{(2\pi)^3} \ \dfrac{e^{-i \bvec q \cdot (\bvec x- \bvec y)}}{-|\bvec{q}|^2 + \qt^2   + i \epsilon} \;,
	\label{eq:amplitude_no_time}
\end{align}
where $\qt^2 = (p^0)^2 - m_{\Phi}^2$ (with $q^0 = p^0 = k^0$), and  $\xb$ and $\yb$ are two spatial vectors.  By completing the square in the exponents of~\refs{eq:amplitude_no_time}, we can perform the Gaussian integrals over $d^3 \bvec x$ and $d^3 \bvec y$. Ignoring an overall phase factor $e^{i(\vecp \cdot \xb - \veck \cdot \yb)}$, we may recast the amplitude in~\eqref{eq:amplitude_no_time} into the more convenient form:
\begin{align}
	T_{\rm L}(p,k; \vecl,\dx,\dy)\, =&\ - 2\pi\, \delta(p^0 - k^0) \ \lambda \, g \ \pi^3  \, \dx^3 \,  \dy^3\,  \nonumber\\
	&\times\, e^{-\lrb{|\vecp|^2 \dx^2 +|\veck|^2 \dy^2 }/4} \dint \dfrac{d^3 \bvec q}{(2\pi)^3} \ \dfrac{e^{i \bvec q \cdot \vecL\, -\, \bvec{q}^2 \, \dl^2/4 }}{-|\bvec{q}|^2 + \qt^2   + i \epsilon}\ ,
	\label{eq:amplitude_after_x-y}
\end{align}
with $\vecL = \vecl- \frac{i}{2} \big( \vecp\,\dx^2  + \veck\, \dy^2 \,\big)$, $\vecl = \yb -\xb$, and $\dl^2 = \dx^2 + \dy^2$. 

Our next step will be to integrate over the polar and azimuthal coordinates of ${\bf q}$ in~\eqref{eq:amplitude_after_x-y}. This angular integration is done explicitly in 
Appendix~\ref{app:angular}, so here we only use the generic formula derived in~\eqref{eq:angular_q_proof_final}. By virtue of~\eqref{eq:angular_q_proof_final}, the transition amplitude $T_{\rm L}$ becomes
\begin{align}
	T_{\rm L}(p,k; \xb,\yb,\dx,\dy) =&\ -2\pi\, \delta(p^0 - k^0) \ \lambda \, g \ \dfrac{\pi }{2} \dfrac{\dx^3 \,  \dy^3}{2}  \nonumber\\ &\times\, \dfrac{e^{-\lrb{|\vecp|^2 \dx^2 +|\veck|^2 \dy^2 }/4}}{\absL}  
	\dint_0^\infty dq \ \dfrac{q \sin\lrb{q \absL} e^{- \bvec{q}^2\, \dl^2/4 } }{-q^2 + \qt^2 + i \epsilon}  \;,
	\label{eq:amplitude_after_angular}
\end{align}
with $q\equiv |\bvec q| \in \mathbb{R}$ and $\absL \equiv \sqrt{\vecL \cdot \vecL} \in \mathbb{C}$. 


The last step will be to integrate over the radial momentum coordinate $q =|{\bf q}|$ in~\eqref{eq:amplitude_after_angular}. 
As outlined in Appendix~\ref{app:radial}, this last radial integration leads to a result expressed in terms of the well-documented complementary error functions~\cite{abramowitz+stegun},
which is the one given by~\eqref{eq:amplitude_final} and reported earlier in~\cite{Ioannisian:1998ch}. 

Alternatively, to make contact with diffractive optics, we can re-express the amplitude~$T_{\rm L}$ in terms of the well-known Fresnel integrals~\cite{abramowitz+stegun} as
\begin{align}
	T_{\rm L}(p,k; \vecl,\dx,\dy)\, =&\ 2\pi \, \delta(p^0 - k^0)\: \dfrac{\lambda \, g \,\pi^2}{4(1+i)}\,    \dfrac{ \dx^3 \,  \dy^3}{\absL}\;  
	e^{- \lrsb{\lrb{|\vecp|^2+\qt^2} \dx^2 +\lrb{|\veck|^2+\qt^2} \dy^2 }/4}  \nonumber \\  
	&\times \bigg[\!
		e^{i\qt \absL} \ \text{C}\!\lrb{\dfrac{2 z_-}{\sqrt{\pi} (1-i)}}\, -\, e^{-i\qt \absL} \ \text{C}\!\lrb{\dfrac{2 z_+}{\sqrt{\pi} (1-i)}} \nonumber\\           
		&+\, i\,e^{i\qt \absL} \ \text{S}\!\lrb{\dfrac{2 z_-}{\sqrt{\pi} (1-i)}}\, -\,i\,e^{-i\qt \absL} \ \text{S}\!\lrb{\dfrac{2 z_+}{\sqrt{\pi} (1-i)}} \nonumber\\
		&-i(1+i)\,  \sin \lrb{\qt\,\absL}
	\bigg] 
	\;.
	\label{eq:amplitude_fresnel}
\end{align}
Here, the Fresnel integrals are
\begin{eqnarray}
	\text{C}(z) = \int_{0}^{z} \ dt \ \cos\lrb{ \pi t^2 / 2}, \quad \text{S}(z) = \int_{0}^{z} \ dt \ \sin\lrb{ \pi t^2 / 2} \;, 
	\label{eq:fresnel_def}
\end{eqnarray}
which are analytically continued to complex arguments $z\in \mathbb{C}$.

\section{Angular Integration}\label{app:angular}
\setcounter{equation}{0}

To calculate the angular part of the integral in~\refs{eq:amplitude_after_x-y}, we first note that 
\begin{equation}
	\dint d^3 \bvec q \ G(|\bvec q|)  \ e^{i \, \bvec q \cdot \vecL} =  \dint_{0}^\infty d | \bvec q | \ G(|\bvec q|)\, |\bvec q|^{n+2}  \
	\displaystyle\sum_n \dfrac{i^n}{n!} \dint_{0}^{2\pi} d\phi \dint_{-1}^{1} d\cos\theta \ ({\bf\hat{q}} \cdot \vecL )^n \;,
	\label{eq:angular_q_proof_step0}
\end{equation}   
where we assume that $G(|\bvec q|)$ falls off sufficiently quickly, so that the integral converges, allowing for the operations of sum and integration to be exchanged. Since $\vecL = \vecl + i \, \bvec w$ is in general a complex vector, we cannot simply rotate the axes in order to bring simultaneously the vectors~$\vecl$ and~$\bvec w$ on the $z$-$y$ plane. However, we note that this integral should be invariant under $O(3)$ rotations. The latter implies 
\begin{equation}
	\dint_{0}^{2\pi} d\phi \dint_{-1}^{1} d\cos\theta \ ({\bf\hat{q}}  \cdot \vecL)^n 
 \, =\,  2 \pi \, \alpha_{n} \, \absL^n \;,
	\label{eq:angular_q_proof_step1}
\end{equation}   
where the coefficients $\alpha_n$ are constants that do not depend on $\vecL$. Consequently, we may determine these constants by calculating the respective integrals using the real projection of $\vecL$, $\vecl \in \mathbb{R}^3$. Thus, we get  
\begin{equation}
	\dint_{0}^{2\pi} d\phi \dint_{-1}^{1} d\cos\theta \ ({\bf\hat{q}}  \cdot \vecl)^n = 2\pi \, \absl^n \dint_{-1}^{1} dz \, z^n \;.
	\label{eq:angular_q_proof_step2}
\end{equation}   
From the latter, we may deduce $\alpha_{n} = \dint_{-1}^{1} dz \, z^n$. Taking this last relation into consideration, we may evaluate the angular part of the integral in~\eqref{eq:angular_q_proof_step0} as follows:
\begin{align}
   \label{eq:angular_q_proof_final}
	\dint d^3 \bvec q \ G(|\bvec q|)  \ e^{i \, \bvec q \cdot \vecL} 
 =&\ 2 \pi \dint_{0}^\infty d | \bvec q | \, |\bvec q|^{2}\ G(|\bvec q|)\ 
	\displaystyle\sum_n \dfrac{i^n}{n!}  \dint_{-1}^{1} dz \ |\bvec q|^{n} \, \absL^n \, z^n \nonumber \\
	=&\ 2\pi \dint d |\bvec q|\, |\bvec q|^2 \ G(|\bvec q|)\ \dint_{-1}^{1} dz\ e^{i \, |\bvec q| \absL \, z} \\[2mm]
 =&\ \dfrac{4\pi}{\absL} \dint_{0}^{\infty} d |\bvec{q}| \, |\bvec q| \ G(|\bvec{q}|)  \, \sin \lrb{|\bvec q| \, \absL}\;.\nonumber
\end{align}

\section{Radial Integration}\label{app:radial}
\setcounter{equation}{0}

In \eqref{eq:amplitude_after_angular}, we have to evaluate the integral over the radial momentum coordinate $q=|{\bf q}|$. To~do~so, we first rewrite it as follows:
\begin{align}
	I\, =&\  \dint_0^\infty dq  \ \dfrac{q \sin\lrb{q \absL} e^{- \bvec{q}^2 \, \dl^2/4 } }{-q^2 + \qt^2 + i \epsilon}\ =\ 
	-\dfrac{1}{2} \dint_{-\infty}^\infty dq  \ \dfrac{q \sin\lrb{q \absL} e^{- \bvec{q}^2 \, \dl^2/4 } }{q^2 - \qt^2 - i \epsilon} \nonumber \\
	=&\ \dfrac{1}{2} \dfrac{\partial}{\partial \absL }\dint_{-\infty}^\infty dq  \ \dfrac{e^{i \, q \, \absL} \, e^{- \bvec{q}^2 \, \dl^2/4 } }{q^2 - \qt^2 - i \epsilon}\;.
	\label{eq:def_integral_q}
\end{align}
The last expression can be further simplified with the help of the Schwinger representation of the propagator, 
\begin{equation}
	\dfrac{1}{{q^2 - \qt^2 - i \epsilon}} = i \dint_{0}^{\infty} dt \ e^{-it (q^2 - \qt^2 - i \epsilon)}\,.
	\label{eq:Schwinger_proper_time_integral}
\end{equation}
Then, the integral~\refs{eq:def_integral_q} may be rewritten as
\begin{align}
	I\, =&\ \dfrac{i}{2} \dfrac{\partial}{\partial \absL }\dint_{-\infty}^\infty dq \dint_{0}^{\infty} dt \, 
	\ e^{-it (q^2 - \qt^2 - i \epsilon)} \,
	e^{i \, q \, \absL} \, e^{- \bvec{q}^2 \, \dl^2/4 }
 \nonumber \\  
	=&\ - \dfrac{\sqrt{\pi}}{2} L\,e^{-\qt \dl^2/4} \dint_{\dl^2}^{+i\infty}\!\!\!du \, \dfrac{e^{-b^2 \, u - L^2/u}}{u^{3/2}} 
	\;.
	\label{eq:integral_q_after_Schwinger}
\end{align}
In the last equality of~\eqref{eq:integral_q_after_Schwinger}, we have introduced the parameters: $L = |\vecL |$ and $b^2 = -(\qt^2 +i\epsilon)/4$. Furthermore, the contour of the complex integration with respect to $u$ is taken to be along the line: $u(t) =\dl^2 + 4 i \, t$, with  $t\in [0,+\infty)$.

We note that the integrand is analytic on the integration contour. This is to be expected, since this corresponds to a simple change of variables. This means that we only need to find the anti-derivative 
of the integrand, and take the appropriate limits. We do so by first expressing~\refs{eq:integral_q_after_Schwinger} in an equivalent form:
\begin{align}
   \label{eq:integral_q_calculation_part_1}
	I\, =&\ - \dfrac{\sqrt{\pi}}{4} L\, e^{-\qt \dl^2/4} \lrsb{
		e^{2 b L} \dint_{\dl^2}^{+i\infty}\!\!\!du\,  \dfrac{e^{- (b\,u+L)^2/u}}{u^{3/2}}\: +\: 
		e^{-2 b L} \dint_{\dl^2}^{+i\infty}\!\!\!du\,  \dfrac{e^{- (b\,u-L)^2/u}}{u^{3/2}}} \\	
	=&\ - \dfrac{\sqrt{\pi}}{4}  e^{-\qt \dl^2/4} \lrsb{
		- e^{ 2 b L} \dint_{\dl^2}^{+i\infty}\!\!\!du\,  \dfrac{b\,u- L}{u^{3/2}} e^{- (b\,u+L)^2/u}\: +\: 
		e^{-2 b L} \dint_{\dl^2}^{+i\infty}\!\!\!du\,  \dfrac{b\,u+L}{u^{3/2}} e^{- (b\,u-L)^2/u}}
	\;,\nonumber
\end{align}
with $b = \pm \big[i \qt -\epsilon/(2\qt)\big]/2$.

After observing that 
\begin{equation}
   \label{eq:AntiErfc}
-\sqrt{\pi} \: \dfrac{d}{du}\,\Erfc\bigg( \frac{b\,u \pm L}{\sqrt{u}}\bigg)\ =\ \dfrac{b\,u \mp L}{u^{3/2}}\; e^{- (b\,u \pm L)^2/u}\;,
\end{equation}
we may now employ this relation to re-express the integrals in the second equality of~\eqref{eq:integral_q_calculation_part_1} as follows:
\begin{align}
	I=& - \dfrac{\sqrt{\pi}}{4}  e^{-\qt \dl^2/4} \sqrt{\pi} \lrsb{ 
		e^{2bL}\Erfc\lrb{\dfrac{b\,u+L}{\sqrt{u}}} - e^{-2bL}\Erfc\lrb{\dfrac{b\,u-L}{\sqrt{u}}}  
	}_{u=\dl^2}^{u\to+i\infty}
	\;.
	\label{eq:integral_q_calculation_part_2}
\end{align}
To evaluate the limit of this expression at $u \to +i \infty$, we introduce a real positive quantity~$R$ (with $R \gg \dl$), such that 
\begin{equation}
    \sqrt{u}\ =\ \frac{1+i}{\sqrt{2}} \, R\ +\ {\cal O}(\dl )\;. 
    \label{eq:u_to_R}
\end{equation}
Then, $u \to  +i \infty$ implies $R\to +\infty$. Thus, 
$|L/\sqrt{u}| \sim |L|/R  \to 0$. In order to determine the infinite limit that appears in \refs{eq:integral_q_calculation_part_2}, we need to evaluate
\begin{eqnarray}
	\Lambda_{\pm} = \lim\limits_{u \to +i\infty} \Erfc\lrb{ b\,\sqrt u }= \lim\limits_{R \to +\infty} \Erfc\bigg[\pm \frac{1+i}{2\,\sqrt{2}} \, \bigg(i \qt - \frac{\epsilon}{2\qt}\bigg) \, R\,\bigg] \;,
\end{eqnarray}
where the sign $\pm$ depends on our choice of the branch we choose for $b$. We find that $\Lambda_{+} = 2$ and $\Lambda_{-} = 0$. Choosing the negative branch of $b$ for the integral~$I$ in~\refs{eq:integral_q_calculation_part_2} yields
\begin{align}
	I\ =\ -\dfrac{\pi}{4} e^{-\qt^2 \dl^2/4} \lrsb{
		e^{i \qt \absL} \, \Erfc\lrb{- \dfrac{i}{2} \qt \, \dl - \dfrac{\absL}{\dl}} - e^{-i \qt \absL} \, \Erfc\lrb{- \dfrac{i}{2} \qt \, \dl + \dfrac{\absL}{\dl}}
	}\;,
	\label{eq:integral_q_result}
\end{align}
after replacing $L$ back with $\absL$. 
It should be pointed out that the same result for the integral~$I$ would be obtained if the positive branch of $b$ was chosen. This follows from the fact that  the difference between the positive ($I_+$) and negative ($I_-$) branch in~\refs{eq:integral_q_calculation_part_2} gives a $u$-independent expression evaluated at two different integration limits, viz.
\begin{equation}
    I_{+} -I_{-}\ \propto\ \sinh\lrsb{ \lrb{i \qt - \dfrac{\epsilon}{2\qt}} L}\bigg|_{u=\dl^2}^{u\to +i\infty} =\ 0 \;.
    \label{eq:branch_diff}
\end{equation}  
Interestingly enough, the above exercise demonstrates that although~\refs{eq:integral_q_after_Schwinger} is symmetric under the exchange of $b \to -b$, maintaining this property of~$I$ in its final expression in~\eqref{eq:integral_q_result} requires a more nuanced treatment.

\newpage
\bibliography{refs}{}
\bibliographystyle{JHEP}                        

\end{document}

%% file: macros.tex
\usepackage{ifthen}
\usepackage{tikz}
\usepackage{xspace}
\usepackage[T1]{fontenc}

\newcommand{\dx}{\delta x\xspace}
\newcommand{\dy}{\delta y\xspace}
\newcommand{\dl}{\delta \ell\xspace}
\newcommand{\xb}{\svev{ \bvec x}\xspace}
\newcommand{\yb}{\svev{ \bvec y}\xspace}

\newcommand{\xbL}{\svev{x}\xspace}
\newcommand{\ybL}{\svev{y}\xspace}

\newcommand{\bvec}[1]{\vectorbold{#1}\xspace}

\newcommand{\vecL}{\bvec L\xspace}
\newcommand{\veck}{\bvec k\xspace}
\newcommand{\vecp}{\bvec p\xspace}
\newcommand{\hatp}{\bvec{\hat p}\xspace}

\newcommand{\vecl}{\pmb{\ell}\xspace}
\newcommand{\hatl}{\pmb{\hat \ell}\xspace}

\newcommand{\absL}{|\vecL|\xspace}
\newcommand{\qt}{\tilde{q}\xspace}
\newcommand{\pmax}{|\vecp|_{\rm *}\xspace}
\newcommand{\absk}{|\veck|\xspace}
\newcommand{\absp}{|\vecp|\xspace}
\newcommand{\absl}{|\vecl|\xspace}

\newcommand{\Erfc}{{\rm Erfc}\xspace}

\newcommand{\dint}{\displaystyle\int \xspace}

\newcommand{\ie}{{i.e.}\xspace}
\newcommand{\eg}{{e.g.}\xspace}
\newcommand{\GeV}{{\rm GeV}\xspace}

\newcommand{\svev}[1]{\langle #1 \rangle}

\newcommand{\lrb}[1]{\left( #1 \right)}
\newcommand{\lrsb}[1]{\left[ #1 \right]}

\newcounter{NumArgs}

\newcommand{\eqs}[1]{\setcounter{NumArgs}{0}\foreach\i in{#1}{\stepcounter{NumArgs}}%
\ifthenelse{\equal{\theNumArgs}{1}}{eq.~(\ref{#1})}%
{\ifthenelse{\equal{\theNumArgs}{2}}%
{eqs.~\foreach\i[count=\q]in{#1}{\ifthenelse{\equal{\q}{\theNumArgs}}{and (\ref{\i})}{(\ref{\i})~}}}%
{eqs.~\foreach\i[count=\q]in{#1}{\ifthenelse{\equal{\q}{\theNumArgs}}{and (\ref{\i})}{(\ref{\i}),~}}}}}

\newcommand{\Eqs}[1]{\setcounter{NumArgs}{0}\foreach\i in{#1}{\stepcounter{NumArgs}}%
\ifthenelse{\equal{\theNumArgs}{1}}{Eq.~(\ref{#1})}%
{\ifthenelse{\equal{\theNumArgs}{2}}%
{Eqs.~\foreach\i[count=\q]in{#1}{\ifthenelse{\equal{\q}{\theNumArgs}}{and (\ref{\i})}{(\ref{\i})~}}}%
{Eqs.~\foreach\i[count=\q]in{#1}{\ifthenelse{\equal{\q}{\theNumArgs}}{and (\ref{\i})}{(\ref{\i}),~}}}}}

\newcommand{\refs}[1]{\setcounter{NumArgs}{0}\foreach\i in{#1}{\stepcounter{NumArgs}}%
\ifthenelse{\equal{\theNumArgs}{1}}{(\ref{#1})}%
{\ifthenelse{\equal{\theNumArgs}{2}}%
{\foreach\i[count=\q]in{#1}{\ifthenelse{\equal{\q}{\theNumArgs}}{and (\ref{\i})}{(\ref{\i})~}}}%
{\foreach\i[count=\q]in{#1}{\ifthenelse{\equal{\q}{\theNumArgs}}{and (\ref{\i})}{(\ref{\i}),~}}}}}

\newcommand{\Figs}[1]{\setcounter{NumArgs}{0}\foreach\i in{#1}{\stepcounter{NumArgs}}%
\ifthenelse{\equal{\theNumArgs}{1}}{Figure~\ref{#1}}%
{\ifthenelse{\equal{\theNumArgs}{2}}%
{Figures~\foreach\i[count=\q]in{#1}{\ifthenelse{\equal{\q}{\theNumArgs}}{and \ref{\i}}{\ref{\i}~}}}%
{Figures~\foreach\i[count=\q]in{#1}{\ifthenelse{\equal{\q}{\theNumArgs}}{and \ref{\i}}{\ref{\i},~}}}}}

\newcommand{\Gen}[2]{\setcounter{NumArgs}{0}\foreach\i in{#2}{\stepcounter{NumArgs}}%
	\ifthenelse{\equal{\theNumArgs}{1}}{#1.~(\ref{#2})}%
	{\ifthenelse{\equal{\theNumArgs}{2}}%
		{#1.~\foreach\i[count=\q]in{#2}{\ifthenelse{\equal{\q}{\theNumArgs}}{and (\ref{\i})}{(\ref{\i})~}}}%
		{#1.~\foreach\i[count=\q]in{#2}{\ifthenelse{\equal{\q}{\theNumArgs}}{and (\ref{\i})}{(\ref{\i}),~}}}}}
